\documentclass[journal]{IEEEtran} 
\usepackage[utf8]{inputenc} 
\usepackage[T1]{fontenc}

\usepackage[pdftex]{graphicx}

\usepackage{cite}

\usepackage[cmex10]{amsmath}
\let\Gamma\varGamma
\usepackage{amsfonts}
\usepackage{amssymb}
\usepackage{mathrsfs}
\usepackage{siunitx}
\usepackage{ifthen}
\usepackage{mathtools}
\usepackage{relsize}
\usepackage{txfonts}

\usepackage{amsthm}

\emergencystretch=\maxdimen
\usepackage{subcaption}
\usepackage{caption}

\usepackage[dvipsnames]{xcolor}

\usepackage{multirow}
\usepackage{booktabs}
\usepackage{array}

\newcommand{\xth}[1]{{\textit{#1}}\mbox{-th}}

\newtheorem{remark}{Remark}
\renewcommand{\qed}{\hfill\blacksquare}

\begin{document}

\title{An $M$-ary Concentration Shift Keying with Common Detection Thresholds For Multitransmitter Molecular Communication}

\author{Ethungshan~Shitiri,~\IEEEmembership{Member,~IEEE,}
        and~Ho-Shin~Cho,~\IEEEmembership{Senior Member,~IEEE}
\thanks{E. Shitiri is with the Department of Computer Architecture, Universitat Polit\`ecnica de Catalunya, Barcelona, Spain, and H.S. Cho is with the School
of Electronic and Electrical Engineering, Kyungpook National University, Daegu, South Korea. 
e-mail: hscho@ee.knu.ac.kr.

This study was supported by the National Research Foundation of Korea (NRF) grant funded by the Korean Government (MSIT) (2021R1A2C1003507).}
\thanks{Manuscript received XXX, 2022; revised XXX, 2022.}

}

\markboth{DRAFT}%
{Shell \MakeLowercase{\textit{et al.}}: Bare Demo of IEEEtran.cls for IEEE Journals}

\maketitle

\begin{abstract}
Concentration shift keying (CSK) is a widely studied modulation technique for molecular communication-based nanonetworks, which is a key enabler for the Internet of Bio-NanoThings (IoBNT). Existing CSK methods, while offering optimal error performance, suffer from increased operational complexity that scales poorly as the number of transmitters, $K$, grows. In this study, a novel $M$-ary CSK method is proposed: CSK with common detection thresholds (CSK-CT). CSK-CT uses \textit{common} thresholds, set sufficiently low to guarantee the reliable detection of symbols from all transmitters, regardless of distance. Closed-form expressions are derived to obtain the common thresholds and release concentrations. To further enhance error performance, the release concentration is optimized using a scaling exponent that also optimizes the common thresholds. The performance of CSK-CT is evaluated against the benchmark CSK across various $K$ and $M$ values. CSK-CT has an error probability between $10^{-7}$ and $10^{-4}$, which is a substantial improvement from that of the benchmark CSK (from $10^{-4}$ to $10^{-3}$). In terms of complexity, CSK-CT is $\mathcal{O}\big(n\big)$ and does not scale with $K$ but $M$ ($M\ll K$), whereas the benchmark is $\mathcal{O}\big(n^2\big)$. Furthermore, CSK-CT can mitigate inter-symbol interference (ISI), although this facet merits further investigation. Owing to its low error rates, improved scalability, reduced complexity, and potential ISI mitigation features, CSK-CT is particularly advantageous for IoBNT applications focused on data gathering. Its effectiveness is especially notable in scenarios where a computationally limited receiver is tasked with collecting vital health data from multiple transmitters.  
\end{abstract}


\IEEEpeerreviewmaketitle

\section{Introduction} \label{sec_intro}

\IEEEPARstart{T}{he} latest addition to the Internet of Things family is the Internet of Bio-Nano Things (IoBNT), enabled by advancements in synthetic biology and nanotechnology \cite{Akyildiz2015IoBnT}. Architecturally, IoBNT comprises three networks at different length scales---the Internet at the macro-scale, wireless body area networks at the micro-scale, and the emerging \textit{nanonetwork} at the nano-scale \cite{Li2022IoNTInterface}. A nanonetwork, comprising interconnected \textit{bio-nanothings} based on living cells and their biochemical functions, can perform basic sensing, communication, and computing tasks \cite{akyildiz2008nanonetworks}. Nanonetworks, with their size advantage, are poised to revolutionize a range of wireless applications, from intra-body healthcare to
hazardous environments in chemical industries \cite{Felicetti2016Applications, Akkas2020Healthcare,Shitiri2021Timing,Al-Zubi2022Intrabody}. However, their small scale presents unique challenges, including strict limitations on energy, computing power, and processing capacity. These constraints make it difficult to
apply conventional, yet complex, wireless technologies, which necessitates the development of simpler, efficient communication methods. 

To this end, molecular communication (MC) is the most promising approach for nanonetworks that draws inspiration from nature and uses engineered molecules to carry information symbols \cite{nakano2005molecular}. MC is suited for nanonetworks due to its nano-scale nature, energy efficiency, and compatibility with biological systems \cite{Chude2017SurveyMCTDDD, Guo2021, Veletic2022Omplants}. Although still a developing technology, it is proving vital for IoBNT's potential healthcare innovation, such as smart drug delivery systems and early disease detection systems \cite{Chude2017SurveyMCTDDD, Reza2019EarlyCancer,Akyl2020Panacea}. This underscores the importance of focused research in this emerging field and is therefore the focus of this study. 

Owing to the inherently stochastic nature of MC and the aforementioned rigid restrictions, conventional wireless techniques are not directly applicable to MC-based nanonetworks \cite{Akyildiz2015IoBnT,nakano2005molecular}. Thus, developing novel modulation techniques
has become a key research focus. Various innovative methods are being explored, such as concentration \cite{Mahfuz2010characterization, Mahfuz2014Comprehensive, Singhal2015Performance, Jamali2018Noncoherent,Wang2023EffConsteCSK}, type\cite{Kim2012Isomers, Kabir2015DMoSK, Wang2020PerforAnalyD-MoSK, Tang2021MolTypePerm}, timing \cite{Srinivas2012TimeMod,Aeeneh2020TimingModulation,Li2020Time-BasedModulation}, direction \cite{Aghababaiyan2020DSK, Aghababaiyan2022BDSK}, and media \cite{Brand2023SwitchingMolecules}. Among these, concentration modulation, or concentration-shift keying (CSK), is widely used because of its simplicity.

CSK is analogous to the conventional amplitude shift keying method, where information symbols modulate the number of molecules released \cite{Mahfuz2010characterization, Mahfuz2014Comprehensive}; CSK operation is described in Section \ref{sec_RelatedWorks}. For optimal error performance, conventional estimation techniques are used, such as the robust maximum likelihood estimation (MLE) \cite{Singhal2015Performance, Jamali2018Noncoherent,Wang2023EffConsteCSK}. These approaches assume that bio-nanothings have adequate computational resources, which is supported by the recent advances in nano-scale computing bio-hardware, such as DNA-based computing \cite{Song2019DNAlogiccircuits, Wang2020DNAswitchingcircuits}. However, the practicality of implementing computationally intensive techniques, such as MLE, on such bio-hardware is not fully understood. Furthermore, integrating computing and communication hardware blocks is not well understood and is an open research challenge \cite{Liu2022DNAComputingMC}. 

Despite advances in computing bio-hardware, low complexity and high energy efficiency must be ensured. Current CSK methods require $K\cdot(M-1)$ threshold computations, where $K$ and $M$ denote the number of transmitters and information symbols, respectively. This results in a quadratic time complexity of $\mathcal{O}\big(n^2\big)$, where $n$ is a function of $K$ and $M$; further details on the calculation of complexity are discussed in Section \ref{sec_timeComplex}. Thus, the complexity scales poorly when $K$ increases in the order of thousands, which is a practical scenario in nanonetworks \cite{akyildiz2008nanonetworks}. Therefore, for CSK methods to be effective, both operational complexity and error performance must be simultaneously addressed. This will help reduce the gap between theoretical CSK frameworks and their practical implementations.

To address the aforementioned problems, a novel low-complexity method, termed CSK with common detection thresholds (CSK-CT), is proposed herein. ``\textit{Common detection threshold}'' reflects CSK-CT's unique approach of computing a single \textit{common} set of $(M-1)$ thresholds, which is valid for every transmitter, unlike current methods that require distinct threshold sets for each transmitter. The common thresholds are sufficiently low to reliably detect symbols transmitted by distant transmitters, which may have weaker received signal strengths. CSK-CT exhibits a linear time complexity, i.e., $\mathcal{O}\big(n\big)$, where $n$ is a function of $M$ and, therefore, does not scale with $K$. Importantly, such complexity makes it on par with the current methods for single transmitter systems and superior for multiple transmitter systems. 

The main contributions of this study are:
\begin{itemize}
    \item Using common thresholds instead of individual thresholds for all different-distance transmitters, resulting in improved time complexity. 
    \item Deriving the closed-form expressions for determining the thresholds and release concentrations simultaneously that are, to the best of our knowledge, demonstrated for the first time. Although not shown, the expressions can hold for different diffusion coefficients or the radius of the receiver or in the presence of flow, enabling the exploration of various physical and network designs. 
    \item Analyzing the correlation between the network parameters (e.g., distance) and CSK-CT parameters (e.g., threshold and release concentration) to reveal how design choices affect performance. This knowledge can be leveraged to design MC-based nanonetworks considering the allowable error probability.
\end{itemize}
   
The remainder of this paper is organized as follows: Section \ref{sec_RelatedWorks} discusses the background on CSK. Section \ref{sec_systemModel} presents the MC free-diffusion channel model, outlining the channel impulse response (CIR), received signal strength, and the average inter-symbol interference (ISI) model. Section \ref{sec_Proposed} discusses the benchmark CSK \cite{Singhal2015Performance}, highlighting the limitations of current CSK methods before introducing the proposed CSK-CT method. Specifically, this section derives the closed-form expressions for calculating threshold values and release concentrations with a scaling exponent. Time complexity and symbol error probability are analyzed in Section \ref{sec_PerfAnaly}. In Section \ref{sec_Results}, the performance of CSK-CT is analyzed for varying values of $M$ (BCSK ($M=2$) and 4-CSK ($M=4$)), average distance, scaling exponent, and symbol period duration. Additionally, the performance of CSK-CT is compared with the benchmark CSK. Section \ref{sec_conclusions} summarizes the study and discusses potential directions for future research.

\section{Background} \label{sec_RelatedWorks}
In CSK, the number of molecules transmitted by a transmitter (a bio-nanothing) is termed release concentration, denoted as $Q$. In general, $M$ release concentration levels represent $M=2^b$ symbols, where $b$ is the number of bits per symbol. The \xth{j} information symbol is denoted as $S_\mathit{\!j}$, and its associated release concentration is denoted as $Q_\mathit{j}$. At the receiver (also a bio-nanothing), the number of molecules received is compared against the detection thresholds, denoted as $\tau$, to decode the transmitted symbol. The total number of molecules received is defined as the received signal strength, denoted as $\mathcal{E}_\text{tot}$. 

\begin{figure}[!t] 
        \centering        \includegraphics[width=0.9\columnwidth,keepaspectratio]{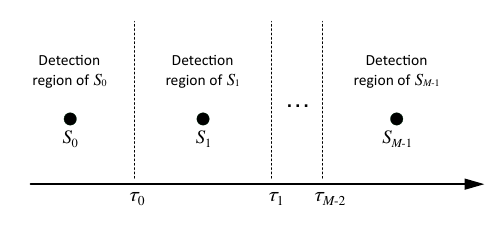}
        \caption {\label{fig_illustrationOfThresholds} Threshold levels and detection regions of information symbols in CSK. }
    \end{figure}

As shown in Fig. \ref{fig_illustrationOfThresholds}, $(M-1)$ detection thresholds are required to identify $M$ symbols. The threshold between adjacent symbols $S_\mathit{\!j}$ and $S_{\mathit{\!j}+1}$ is denoted by $\tau_\mathit{j}, j\in[0,M-2]$. Accordingly, the detection region of a symbol $S_\mathit{\!j}$ lies between thresholds $\tau_{\mathit{j}-1}$ and $\tau_\mathit{j}$. Suppose $S_{\!1}$ is transmitted. If $\mathcal{E}_\text{tot}$ falls between $\tau_0$ and $\tau_1$, the symbol is decoded as $S_{\!1}$. If $\mathcal{E}_\text{tot}$ is between $\tau_1$ and $\tau_2$, the symbol is decoded as $S_{\!2}$, and so on.  

Given the infancy and complex nature of MC systems, performance analysis of CSK methods often involves a simplified communication system model, comprising a single transmitter and receiver \cite{Mahfuz2010characterization, Mahfuz2014Comprehensive, Singhal2015Performance, Jamali2018Noncoherent,jamali2019channelTut}. The simplified model has laid the foundation for the fundamental principles of MC. For example, the channel impulse response (CIR) function, which characterizes a communication channel’s behavior in response to an impulse input, can be derived \cite{jamali2019channelTut}. Most importantly, CIR allows the derivation of $\mathcal{E}_\text{tot}$. 

For CSK methods, the characterization of $\mathcal{E}_\text{tot}$ is crucial in assessing system performance, in particular the symbol error probability (or rate), whose minimization is pivotal in assessing the efficiency of a modulation technique \cite{Kuran2021SurveyMod}. Moreover, MC channels have memory \cite{jamali2019channelTut}, implying that molecules from a previous symbol transmission may arrive during a current transmission, leading to ISI \cite{Mahfuz2011ISI}. Since ISI molecules cannot be distinguished from the signal molecules and are incorrectly considered part of the current transmission, they worsen the symbol error probability. 

Under the simplified system model, the number of thresholds to be computed is $M-1$ \cite{Singhal2015Performance}. If the CSK approaches that are developed based on the simplified model are extended to multiple transmitters, as will be discussed in Section \ref{sec_Proposed}, the receiver must compute $K\cdot(M-1)$ thresholds, which increases proportionately with $K$. We emphasize that the high computational requirement of CSK that scales with $K$ could limit its practical application; thus, CSK-CT.  

    \begin{figure}[!t] 
        \centering 
        \includegraphics[width=0.9\columnwidth,keepaspectratio]{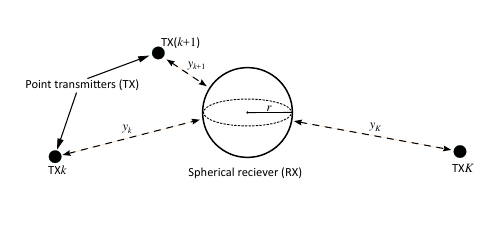}
        \caption {\label{fig_network_topo} Network topology. 
        }
    \end{figure}

\section{System Model} \label{sec_systemModel}
A star-network topology in a three-dimensional (3D) unbounded medium, with $K$ point transmitters randomly distributed around a fully absorbing spherical-shaped receiver with radius $r$ is considered herein (Fig. \ref{fig_network_topo}). The random distance between any given transmitter and the nearest surface of the receiver surface, denoted by $y$, follows the uniform distribution, i.e., $Y\sim \mathcal{U}(y_\text{min},y_\text{max})$, where $y_\text{min}$ and $y_\text{max}$ represent the minimum and maximum distances in the network, respectively. $\mathcal{U}(a,b)$ represents the continuous uniform distribution with $a$ and $b$ as the minimum and maximum values, respectively. Accordingly, the probability density function (PDF) of $y$ is $f_Y(y)=1/(y_\text{max}-y_\text{min})$. 

\begin{figure}[!t] 
        \centering 
        \includegraphics[width=.99\columnwidth,keepaspectratio]{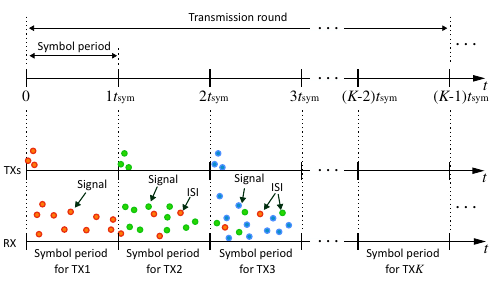}
        \caption {\label{fig_timeslot_model} Time-slotted channel model (top) and time-divided channel access (bottom) highlighting the signal and ISI molecules. 
        }
    \end{figure}

The transmitters are time-synchronized with the receiver and access the channel in a time-slotted manner, similar to time-division multiple access \cite{Shitiri2021Timing, Lin2016ClockSync,Shitiri2021TDMA}. As shown in Fig. \ref{fig_timeslot_model}, the time is divided into equal, recurring intervals called symbol periods, with $t_{\text{sym}}$ denoting the symbol period duration. In each symbol period, a symbol is transmitted at the beginning, and its reception is expected within the same period. Each transmitter is allocated one symbol period in each transmission round. Hence, each transmitter knows when to transmit without collision \cite{Shitiri2021TDMA}. 

The channel is assumed to have a memory of length $(K-1)$. This means the signal received in any symbol period includes the current transmission’s intended signal and ISI from the previous $(K-1)$ transmissions. Thus, each transmitted symbol potentially contributes to ISI in the subsequent $(K-1)$ symbols. For example, in Fig. \ref{fig_timeslot_model}, transmitters TX1, TX2, and TX3 transmit sequentially in the first, second, and third symbol periods. In the first period, the receiver receives only the intended signal of TX1 (red color) with no ISI component due to the absence of prior transmissions. In the second period, the receiver receives the intended signal of TX2 (green color) along with TX1's ISI. Similarly, in the third period, the receiver receives the intended signal of TX3 (blue color) and the ISI from both TX1 and TX2. The details of the ISI formulation are provided in Section \ref{sec_ISI_model}.
    
    
\subsection{Free-diffusion 3D channel model and CIR}
In a diffusive channel, the molecules released by a transmitter move independently of each other and are governed by the laws of Brownian motion \cite{einstein1956BM}. When a molecule hits the surface of the receiver, it gets absorbed. For a fully absorbing receiver, any point on its surface acts as an absorption point \cite{jamali2019channelTut}. As the absorbed molecules are effectively removed from the channel, they contribute to the received signal only once \cite{yilmaz2014Three,Kuscu2019TxRxArchitectures}.
If a transmitter located at a distance $y$ from the receiver releases the molecules at time $t=0$, then the hitting rate of the molecules at the receiver at time $t>0$ is given by \cite{redner2001AGuide}
        \begin{equation} \label{eq_fhitt}
                f_\text{hit}(y,t) = \frac{r}{y+r} \frac{y}{\sqrt{4 \pi D t^3}} e^{-{y}^2/{4Dt}},
        \end{equation}
where $D$ is the diffusion coefficient. The fraction of molecules reaching the receiver by time $t$ can be derived from \eqref{eq_fhitt} as follows \cite{yilmaz2014Three}        
        \begin{equation} \label{eq_Fhitt}
             F_\text{hit}(y,t) = \frac{r}{y+r}  \text{erfc}\left(\frac{y}{\sqrt{4Dt}}\right),
        \end{equation}
where erfc$(\cdot)$ is the complementary error function. The CIR represents the probability of molecules transmitted at time $t=0$ being absorbed by the receiver at a distance $y$ during the \xth{i} symbol period. It is expressed as \cite{jamali2019channelTut} 
\begin{align}\label{eq_CIR}
    \begin{split}
    h(y,i) &=F_\text{hit}\left(y,it_\text{sym}\right) -  F_\text{hit}\left(y,\left(i-1\right)t_\text{sym}\right)
    \end{split}    
\end{align}
By definition, $h(y,1)$ represents the CIR of the signal, and $h(y,i),\ i\ge2$, represents the CIR of the ISI. Referring to Fig. \ref{fig_timeslot_model}, consider TX3 is currently transmitting. In this case, $h(y_\text{3},1)$ denotes the CIR of TX3's signal. Then, $h(y_\text{1},3)$ and $h(y_\text{2},2)$ denote the CIR of the ISI originating from TX1 and TX2, respectively.

\subsection{Received signal strength without ISI and noise}    
Let the symbol $S_\mathit{\!j}\ (j=0,1,2,\ldots,M-1)$ be assigned the release concentration $Q_\mathit{j}$ and be equally transmitted with a probability of $1/M$. Furthermore, $Q_{j+1} > Q_j, \forall j$. Then, the received signal strength, when a transmitter with distance $y$ releases the concentration $Q$, can be expressed as

\allowdisplaybreaks
        \begin{equation} \label{eq_Nrx_SignalPre}
             \mathcal{E}_{\text{sig}|Q} = Q\, h(y,1). 
        \end{equation}
As $Q$ and $y$ are independent random variables, the expected received signal strength can be expressed as 
        \begin{align} \label{eq_Nrx_Signal}
            \mathcal{E}_\text{sig} &= \mathbb{E}\left[ \mathcal{E}_{\text{sig}|Q}\right] \notag \\
             &= \mathbb{E}\left[Q\, h(y,1)\right] \notag \\
             &=\int_{y_\text{min}}^{y_\text{max}} \sum_{Q=Q_0}^{Q_{M-1}} Q\,h(y,1)\, \text{Pr}(Q)\,f_Y(y)\,dy  \notag \\
             &= \sum_{Q=Q_0}^{Q_{M-1}} Q\, \text{Pr}(Q)\,\int_{y_\text{min}}^{y_\text{max}}\,h(y,1)\,f_Y(y)\,dy  \notag \\
             &= \frac{Q_0+\ldots+Q_{M-1}}{M}\, \frac{1}{y_\text{max}-y_\text{min}}\int_{y_\text{min}}^{y_\text{max}}\,h(y,1)\,dy \notag \\
             &=\bar{Q}\, \frac{1}{y_\text{max}-y_\text{min}}\int_{y_\text{min}}^{y_\text{max}}\,\left(F_\text{hit}(y,t_\text{sym}) -  F_\text{hit}\left(y,0\right)\right)\,dy \notag
             \\
             &=\bar{Q}\, \frac{1}{y_\text{max}-y_\text{min}}\int_{y_\text{min}}^{y_\text{max}}\,\frac{r}{y+r}  \text{erfc}\,\Bigg(\frac{y}{\sqrt{4Dt_\text{sym}}}\Bigg)\,dy \notag  \\
            &= \bar{Q}\, \mathcal{H}[1],
        \end{align}
where $\mathbb{E}[\cdot]$ is the expectation operator, $\text{Pr}(Q)$ is the probability that $Q$ is transmitted, and $\mathcal{H}[1]=\frac{1}{y_\text{max}-y_\text{min}}$ $\int_{y_\text{min}}^{y_\text{max}}\,\frac{r}{y+r}  \text{erfc}\left(y/\sqrt{4Dt_\text{sym}}\right)\,dy$ is the average CIR of the signal. Note that the integral is intractable and is therefore solved numerically.



\subsection{Received signal strength with ISI and without noise}\label{sec_ISI_model} 
\begin{figure}[!t] 
        \centering 
        \includegraphics[width=0.9\columnwidth,keepaspectratio]{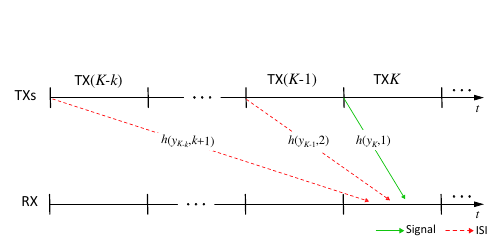}
        \caption {\label{fig_ISI_timeslot_model} CIRs of signal (green solid line) and ISI from previous $k$ transmissions from other transmitters (red dashed lines). 
        }
    \end{figure}

To derive the received signal strength with ISI, the ISI is first characterized by considering at most $k$ previous transmissions $ (1\le k \le K-1)$ as shown in Fig. \ref{fig_ISI_timeslot_model}. Then, the expected ISI is the cumulative interference from each prior transmission, expressed as
\begin{align}\label{eq_Nrx_ISI}
    \mathcal{E}_\text{isi}&= \sum_{i=2}^{k+1}\mathbb{E}\left[Q\ h(y,i)\right]\notag \\
    &= \sum_{i=2}^{k+1}\int_{y_\text{min}}^{y_\text{max}} \sum_{Q=Q_0}^{Q_{M-1}} Q\,h(y,i)\, \text{Pr}(Q)\,f_Y(y)\,dy  \notag \\
    \begin{split}
    &=\bar{Q}\, \sum_{i=2}^{k+1} \frac{1}{y_\text{max}-y_\text{min}}\int_{y_\text{min}}^{y_\text{max}}\,\Bigg(F_\text{hit}\left(y,it_\text{sym}\right) -  \\ &\qquad\qquad\qquad\qquad\qquad\qquad  F_\text{hit}\left(y,(i-1)t_\text{sym}\right)\Bigg)\,dy \notag        
    \end{split}\\
    \begin{split}
    &=\bar{Q}\, \sum_{i=2}^{k+1} \frac{1}{y_\text{max}-y_\text{min}}\int_{y_\text{min}}^{y_\text{max}}\,\frac{r}{y+r}  \Bigg(\text{erfc}\,\Bigg(\frac{y}{\sqrt{4D-it_\text{sym}}}\Bigg) -\\
    & \qquad\qquad\qquad\qquad\qquad\qquad \text{erfc}\,\Bigg(\frac{y}{\sqrt{4D(i-1)t_\text{sym}}}\Bigg)\,\Bigg)\,dy  \notag       
    \end{split}\\
    &= \bar{Q}\sum_{i=2}^{k+1} \mathcal{H}[i],\quad 1\le k \le K-1,
\end{align}
where $\sum_{i=2}^{k+1} \mathcal{H}[i]$ is the sum of the average CIR of the interferes. The integral is intractable and is therefore solved numerically. As the interference from other transmitters is considered, \eqref{eq_Nrx_ISI} also describes the multi-user interference (MUI). However, as one symbol per transmission per transmitter is considered, therefore, without the loss of generality, the term ISI is used instead of MUI. 

Lastly, combining \eqref{eq_Nrx_Signal} and \eqref{eq_Nrx_ISI}, the total expected received signal strength considering ISI can be expressed as 
        \begin{align} \label{eq_Nrx}
        \mathcal{E}_\text{tot} &=\mathcal{E}_\text{sig}+ \mathcal{E}_\text{isi} \notag \\
         &=\bar{Q} \sum_{i=1}^{k+1} \mathcal{H}[i].
        \end{align}

\subsection{Received signal strength with ISI and noise}
Since the released molecules move independently of each other, the received signal strength during the \xth{i} symbol period is random. This can be modeled by the Binomial random variable with $Q$ trials and a success probability $h(y,i)$ \cite{Kuran2010Energy}. When the success probability is sufficiently small, $Q$ is large, and $\Delta t\rightarrow 0$, by mathematical induction, the received signal strength is a random variable following the Normal distribution.

Differing from the general $\bar{Q}$ that is considered in \eqref{eq_Nrx_Signal}, we consider the expectation of the total received signal strength conditioned on any specific $Q_\mathit{j}$. This conditional strength can be expressed as
\begin{align} \label{eq_Nrx_Qm}
    \mathcal{E}_{\text{tot}|Q_\mathit{j}}&= \mathcal{E}_{\text{sig}|Q_\mathit{j}} + \mathcal{E}_\text{isi}. \notag \\
    &=Q_\mathit{j} \mathcal{H}[1] +  \bar{Q}\sum_{i=2}^{k+1} \mathcal{H}[i].
\end{align}
Then, the received signal strength, described by the Normal distribution, is
        \begin{equation} \label{eq_NormalApprox}
            \hat{\mathcal{E}}_{\text{tot}|Q_\mathit{j}}\sim  \mathscr{N}\left(\mathcal{E}_{\text{tot}|Q_\mathit{j}},\, \sigma^2_{\text{tot}|Q_\mathit{j}}\right),
        \end{equation}
where $\mathscr{N}\left(a,b\right)$ represents the Normal distribution with mean $a$ and variance $b$. The variance $\sigma^2_{\text{tot}|Q_\mathit{j}}$ in \eqref{eq_NormalApprox} considers both the inherent noise inherent from Brownian motion \cite{Hoda2013BlindSynchronization} and interference. As the received signal and interference are independent of each other, the variance can be expressed as 
\begin{equation}
    \sigma^2_{\text{tot}|Q_\mathit{j}} = \sigma^2_{\text{sig}|Q_\mathit{j}} +\sigma^2_\text{isi}, 
\end{equation}
where
\begin{subequations}
\begin{align}    \sigma^2_{\text{sig}|Q_\mathit{j}}&=Q_\mathit{j}\, \mathcal{H}[1] \cdot\left(1- \mathcal{H}[1]\right),  \\ \sigma^2_\text{isi}&=  \bar{Q} \sum_{i=2}^{k+1} \mathcal{H}[i] \cdot\left(1-\mathcal{H}[i]\right),\ 1\le k \le K-1.  \label{eq_ISI_var}
\end{align}
\end{subequations}

\section{Proposed CSK with Common Detection Thresholds} \label{sec_Proposed}

Before introducing the proposed CSK-CT method, the generalization of Singhal \textit{et al.}'s CSK method \cite{Singhal2015Performance} to multiple transmitter scenarios is discussed.  Singhal \textit{et al.}'s method is referred to as the benchmark CSK, taking into account its MLE approach.

\subsection{Generalization of the benchmark CSK}
For any given transmitter, denoted by $k$ and located at a random distance $y_k$ from the receiver, the threshold value can be calculated as
\begin{equation}\label{eq_ThresholdSinghal}
    \tau_{k,j} = \sqrt{\mathcal{E}_{\text{tot}|Q_\mathit{j}, y_k}\, \mathcal{E}_{\text{tot}|Q_{\mathit{j}+1},y_k}},\quad \forall k,
\end{equation}
where 
\begin{subequations}\label{eq_Nrx_conditioned}
  \begin{align}
    \mathcal{E}_{\text{tot}|Q_\mathit{j},y_k}&=Q_\mathit{j}\,h(y_k,1)+\mathcal{E}_\text{isi}, \\ \mathcal{E}_{\text{tot}|Q_{\mathit{j}+1},y_k}&=Q_{\mathit{j}+1}\,h(y_k,1)+\mathcal{E}_\text{isi},    
  \end{align}  
\end{subequations} 
and $\tau_{k,j}$ represents threshold $j$ between symbols $S_\mathit{\!j}$ and $S_{\mathit{\!j}+1}$ of transmitter $k$. As indicated in \eqref{eq_ThresholdSinghal}, the receiver needs to obtain $(M-1)$ thresholds for each transmitter, resulting in a total of $K\cdot(M-1)$ thresholds. This also implies that $K$ CIR information is required. 

\subsection{Description of the proposed CSK-CT}
CSK-CT is designed based on the following principles. The \xth{j} threshold value must be sufficiently low to ensure the received signal strength of symbol $S_{\mathit{j}+1}$ from any transmitter can cross it. Conversely, it must be sufficiently large to prevent the received signal strength of symbol $S_\mathit{j}$ from any transmitter from crossing it. Formally, the \xth{j} threshold is \textit{equal to} or \textit{smaller than} the weakest received signal strength that is associated with $Q_{j+1}$. Simultaneously, it is \textit{equal to} or \textit{greater than} the strongest received signal strength that is associated with $Q_j$. These principles can be expressed as
    \begin{equation} \label{eq_ini_ConditionCSK-CT}
        \underset{k}{\operatorname{argmax}} \,\mathcal{E}_{\text{tot}|Q_\mathit{j},y_k}\le\tau_\mathit{j}\le\underset{k}{\operatorname{argmin}}\,\mathcal{E}_{\text{tot}|Q_{\mathit{j}+1}, y_k},\ \forall k \in K,
    \end{equation}
which establishes the lower and upper bounds of $\tau_\mathit{j}$. These bounds can be further simplified. Let $y_\text{min}$ and $y_\text{max}$ denote the minimum and maximum $y$. If $K$ is sufficiently large, the following approximations hold: 
\begin{subequations}\label{eq_approx}
\begin{align}
 \underset{k}{\operatorname{argmax}} \,\mathcal{E}_{\text{tot}|Q_\mathit{j},y_k} &\approx \mathcal{E}_{\text{tot}|Q_\mathit{j},y_\text{min}}, \\
 \underset{k}{\operatorname{argmin}} \,\mathcal{E}_{\text{tot}|Q_{\mathit{j}+1},y_k}&\approx \mathcal{E}_{\text{tot}|Q_{\mathit{j}+1},y_\text{max}},   
\end{align}  
\end{subequations}
 where $\mathcal{E}_{\text{tot}|Q_\mathit{j},y_\text{min}}$ and $\mathcal{E}_{\text{tot}|Q_{\mathit{j}+1},y_\text{max}}$ denote the strongest and weakest received signal strength associated with $Q_j$ and $Q_{j+1}$ at $y_\text{min}$ and $y_\text{max}$, respectively. For clarity, $\mathcal{E}_{\text{tot}|Q_\mathit{j},y_\text{min}}$ and $\mathcal{E}_{\text{tot}|Q_{\mathit{j}+1},y_\text{max}}$ are referred to as the \textit{limits of received signal strength} because they are associated with the distance limits, i.e., $y_\text{min}$ and $y_\text{max}$, respectively. Hence, \eqref{eq_ini_ConditionCSK-CT} simplifies to 
    \begin{equation}  \label{eq_ConditionCSK-CT}
        \mathcal{E}_{\text{tot}|Q_\mathit{j},y_\text{min}}\le\tau_\mathit{j}\le\mathcal{E}_{\text{tot}|Q_{\mathit{j}+1},y_\text{max}},
    \end{equation}
where
\begin{subequations}\label{eq_Nrx_ymin_ymax}
  \begin{align}
    \mathcal{E}_{\text{tot}|Q_\mathit{j},y_\text{min}}&=Q_\mathit{j}\,h(y_\text{min},1)+\mathcal{E}_\text{isi}, \\ \mathcal{E}_{\text{tot}|Q_{\mathit{j}+1},y_\text{max}}&=Q_{\mathit{j}+1}\,h(y_\text{max},1)+\mathcal{E}_\text{isi}.    
  \end{align}  
\end{subequations} 
The release concentrations are therefore strictly ordered as follows:
\begin{subequations} \label{eq_Qcondition}
\begin{align}
    &Q_0<Q_1<\dots<Q_{M-1}, \label{eq_QCondi}\\
    \text{s.t.}&\notag \\ 
    &\mathcal{E}_{\text{tot}|Q_{0},y_\text{min}}\le\mathcal{E}_{\text{tot}|Q_{1},y_\text{max}},\ \mathcal{E}_{\text{tot}|Q_{1},y_\text{min}}\le\mathcal{E}_{\text{tot}|Q_{2},y_\text{max}},\ldots\notag\\
    &\qquad\qquad\qquad
    \mathcal{E}_{\text{tot}|Q_{M-2},y_\text{min}}\le\mathcal{E}_{\text{tot}|Q_{M-1},y_\text{max}}. \label{eq_meanCondi} 
\end{align}
\end{subequations}

As evidenced from \eqref{eq_ConditionCSK-CT} to \eqref{eq_Qcondition}, the $Q$ values must be first derived to obtain the thresholds, but they must fulfill both \eqref{eq_QCondi} and \eqref{eq_meanCondi}. Existing methods often use distance-independent $Q$ values for single transmitter scenarios \cite{Mahfuz2010characterization, Mahfuz2014Comprehensive, Singhal2015Performance, Jamali2018Noncoherent}. However, in CSK-CT's context for multiple transmitter scenarios, using such distance-independent $Q$ values might not satisfy \eqref{eq_meanCondi} even if they meet \eqref{eq_QCondi}. Therefore, a closed-form expression is derived to determine the values of $Q$ that fulfill \eqref{eq_meanCondi}, as discussed below.

\subsubsection{Determining the release concentrations}

For simplicity, we first relax the inequality conditions in \eqref{eq_ConditionCSK-CT}. Then, the following expression can be obtained: 
    \begin{align}\label{eq_tauCond}
         &\tau_\mathit{j} = \mathcal{E}_{\text{tot}|Q_{\mathit{j}+1},y_\text{max}} =\mathcal{E}_{\text{tot}|Q_\mathit{j},y_\text{min}}. 
    \end{align}
Relaxing these conditions does not invalidate \eqref{eq_ConditionCSK-CT}, which defines $\tau_\mathit{j}$ as being at most equal to $\mathcal{E}_{\text{tot}|Q_{\mathit{j}+1},y_\text{max}}$ or at least equal to  $\mathcal{E}_{\text{tot}|Q_\mathit{j},y_\text{min}}$. This approach mitigates 
the intractability posed by the inequality condition. In the latter part of this section, a simplified way (\textit{cf.} \eqref{eq_DetQInequal}) to realize the inequality condition is discussed. 

From \eqref{eq_tauCond}, we note that the received signal strength of symbols $S_{\mathit{j}}$ and $S_{\mathit{j}+1}$ are equal at their respective distance limits. This implies a tight correlation between the release concentrations, $Q_{\mathit{j}}$ and $Q_{\mathit{j}+1}$; therein, $Q_{\mathit{j}+1}$ has to be a function of $Q_{j}$. Then, the solution to determine the values of $Q$ lies in the ratio between the adjacent release concentrations. First, \eqref{eq_Nrx_ymin_ymax} is substituted in \eqref{eq_tauCond} to obtain 
\begin{align}\label{eq_Gamma_relationship}
Q_{\mathit{j}+1}\,h(y_\text{max},1)+\mathcal{E}_\text{isi} &= Q_\mathit{j}\,h(y_\text{min},1)+\mathcal{E}_\text{isi}.
\end{align}
Rearranging \eqref{eq_Gamma_relationship} yields  
       \begin{equation} \label{eq_Gamma_Final}
        \begin{aligned}
         \frac{Q_{\mathit{j}+1}}{Q_\mathit{j}} =          \frac{h(y_\text{min},1)}{h(y_\text{max},1)}  \triangleq\Gamma.
        \end{aligned}
        \end{equation}
Solving \eqref{eq_Gamma_Final} for each $Q_j$, the release concentrations can be derived as
        \begin{equation} \label{eq_DetQ}
           Q_\mathit{j} =  Q_{0}\,\Gamma^{\,j} , 
        \end{equation}
where $Q_0$ is a system parameter. In instances where $Q_\mathit{j}$ is not an integer, the nearest integer to $Q_\mathit{j}$ may be chosen. 

\begin{figure}
\centering
        \includegraphics[width=.9\columnwidth,keepaspectratio]{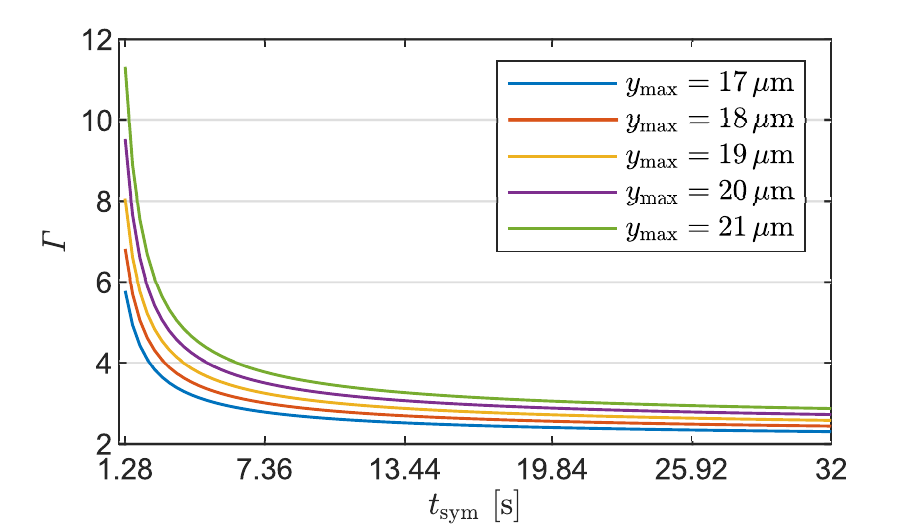}
\caption{$\Gamma$ versus $t_\text{sym}$ for $y_\text{min}=\SI{6}{\micro\metre}$, $D=$ \SI{79.4}{\micro\metre^{2}\per\second}, and $r=$ \SI{5}{\micro\metre}.}
\label{fig_Gamma_Vs_tsym}
\end{figure}

Based on \eqref{eq_Gamma_Final}, the ratio $\Gamma$ is a function of distance, $y_\text{min}$ and $y_\text{max}$, and time, $t_\text{sym}$ (\textit{cf.} \eqref{eq_CIR}). To characterize these relations, in Fig. \ref{fig_Gamma_Vs_tsym} the behavior of $\Gamma$ is analyzed for varying $t_\text{sym}$ and $y_\text{max}$ while keeping $y_\text{min}$ constant. We can observe that $\Gamma$ decreases with $t_\text{sym}$, despite the increase and eventual convergence of $h(y_x,1)$, $x \in \{\text{min},\text{max}\}$, to $\frac{r}{y_x+r}$ \cite{Wang2019ANovel}. The rate of increase of $h(y_\text{min},1)$ is faster compared to $h(y_\text{max},1)$ because $y_\text{min}<y_\text{max}$. This difference indicates that $h(y_\text{max},1)$ still increases when $h(y_\text{min},1)$ converges. Hence, $\Gamma$ decreases with $t_\text{sym}$. Furthermore, $\Gamma$ increases with $y_\text{max}$ because $h(y_\text{max},1)$ decreases with $y_\text{max}$. 

\begin{remark}\label{remark1}
$\Gamma$ is inversely proportional to $t_\text{sym}$ and directly proportional to $y_\text{max}$ when $y_\text{min}$ is constant (or inversely proportional to $y_\text{min}$ when $y_\text{max}$ is constant).  $\qed$
\end{remark}

\begin{figure}
\centering
  \includegraphics[width=.8\columnwidth]{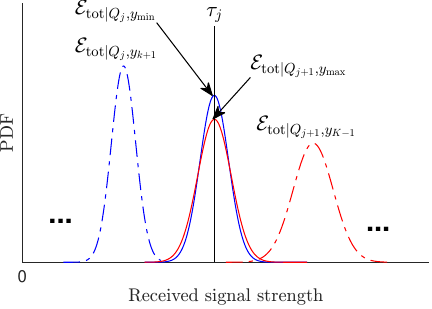}\vspace{0em}
\caption{PDFs of the received signal strength when $\tau_\mathit{j} = \mathcal{E}_{\text{tot}|Q_{\mathit{j}+1},y_\text{max}} =\mathcal{E}_{\text{tot}|Q_\mathit{j},y_\text{min}}$.}
\label{fig_Signal_Spacing_equal}
\end{figure}

Fig. \ref{fig_Signal_Spacing_equal} shows the probability density function of received signal strengths as per \eqref{eq_Nrx_conditioned} and \eqref{eq_Nrx_ymin_ymax}. The blue lines denote the PDFs of $\mathcal{E}_{\text{tot}|Q_{\mathit{j}},y_k}, \forall k$, the red lines denote the PDFs of $\mathcal{E}_{\text{tot}|Q_{\mathit{j}+1},y_k}, \forall k$. Solid lines denote the PDFs of the limits of received signal strength, while dashed lines denote the PDFs at distances other than $y_\text{min}$ and $y_\text{max}$. The PDFs of the limits of received signal strength overlap completely because of the relaxation of inequality. Such overlapping will result in a high error probability at the threshold values; thus, the system's performance deteriorates.

\begin{figure}
 \centering
  \includegraphics[width=.8\columnwidth]{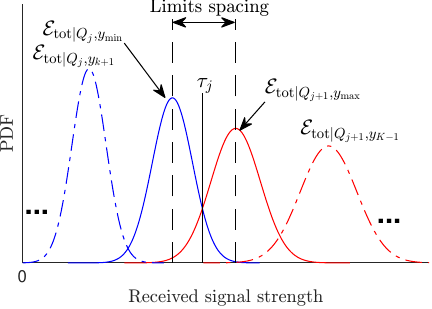}\vspace{0em}
\caption{PDFs of the received signal strength and the limits spacing when $\mathcal{E}_{\text{tot}|Q_\mathit{j},y_\text{min}} <\tau_\mathit{j} < \mathcal{E}_{\text{tot}|Q_{\mathit{j}+1},y_\text{max}}$.}
\label{fig_Signal_Spacing_unequal}
\end{figure}

To reduce the errors, the PDFs of the limits of received signal strength must have minimal or no overlapping at the thresholds. This implies that $\mathcal{E}_{\text{tot}|Q_\mathit{j},y_\text{min}} <\tau_\mathit{j} < \mathcal{E}_{\text{tot}|Q_{\mathit{j}+1},y_\text{max}}$ must hold, which is the inequality condition of \eqref{eq_ConditionCSK-CT}. In other words, there should be a significant space between the limits of received signal strength. This spacing is referred to as the \textit{limits spacing} (Fig. \ref{fig_Signal_Spacing_unequal}).

To ensure the limits spacing is greater than $0$ and thereby reduce errors, adjustments in release concentrations are
necessary. This can be achieved by increasing the release concentrations as it is proportional to $\Gamma$ (\eqref{eq_DetQ}). However, with $y_\text{min}$ and $y_\text{max}$ being constant, $\Gamma$ is fixed and cannot be modified. To circumvent this, an auxiliary parameter is introduced to $\Gamma$ to obtain the \xth{n} power of $\Gamma$, namely, the scaling exponent $\rho$, a positive real number. The release concentration is reformulated as
    \begin{equation} \label{eq_DetQInequal}
           Q_\mathit{j} = Q_{0}\cdot\left(\Gamma^{\, j}\right)^{\,\rho},\ \rho\in \mathbb{R}_{\ge 1}. 
    \end{equation}
Equation \eqref{eq_DetQInequal} is the final expression for calculating the release concentration in CSK-CT.  

\subsubsection{Determining the thresholds}
For cases when $\rho=1$, $\tau_\mathit{j}$ can be calculated using \eqref{eq_tauCond}, but not for those with $\rho>1$ cases. For the latter cases, the method of pairwise PDF comparison, as employed in \cite{Singhal2015Performance}, is applicable for determining $\tau_\mathit{j}$ as $\mathcal{E}_{\text{tot}|Q_\mathit{j},y_\text{min}}< \mathcal{E}_{\text{tot}|Q_{\mathit{j}+1},y_\text{max}},\ \forall j \in M$, must hold. Therefore, for cases with $\rho>1$, $\tau_\mathit{j}$ is given by
    \begin{subequations}
        \begin{align}
          &\tau_\mathit{j} =\sqrt{\mathcal{E}_{\text{tot}|Q_\mathit{j},y_\text{min}}\,\mathcal{E}_{\text{tot}|Q_{\mathit{j}+1},y_\text{max}}}, \label{eq_tauInequality} \\
          \text{s.t.}\ &\mathcal{E}_{\text{tot}|Q_\mathit{j},y_\text{min}}<\mathcal{E}_{\text{tot}|Q_{\mathit{j}+1},y_\text{max}}. \label{eq_MeanInequality} 
        \end{align}
    \end{subequations}
Readers are referred to \cite{Singhal2015Performance} wherein the detailed methodology of deriving \eqref{eq_tauInequality} is discussed. When $\mathcal{E}_{\text{tot}|Q_\mathit{j},y_\text{min}} = \mathcal{E}_{\text{tot}|Q_{\mathit{j}+1},y_\text{max}}$ is applied to \eqref{eq_tauInequality}, it reduces to \eqref{eq_tauCond}. By considering the cases of $\rho$ described by \eqref{eq_tauCond} and \eqref{eq_tauInequality}, the final expression of $\tau_\mathit{j}$ for CSK-CT is
\begin{equation} \label{eq_ThresholdCSK-CT}
    \tau_\mathit{j} = 
\begin{cases}
    \mathcal{E}_{\text{tot}|Q_{\mathit{j}+1},y_\text{max}},& \text{if } \rho= 1 \smash{\raisebox{-1ex}{$\ , \ \forall j.$}} \\
    \sqrt{\mathcal{E}_{\text{tot}|Q_\mathit{j},y_\text{min}}\,\mathcal{E}_{\text{tot}|Q_{\mathit{j}+1},y_\text{max}}},              & \text{if}\ \rho>1
\end{cases}
\end{equation}
Equation \eqref{eq_ThresholdCSK-CT} is the final expression for calculating the thresholds in CSK-CT.

\begin{figure*}
\centering
\begin{subfigure}{1\textwidth}
  \centering
  \includegraphics[clip,width=.85\textwidth]{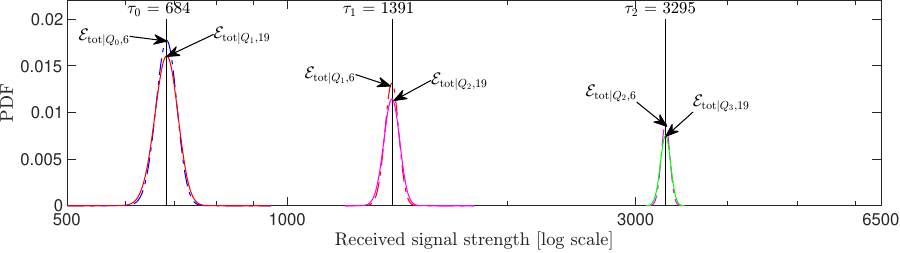}\vspace{-.1em}
  \caption{$\rho=1$} \vspace{1em}
  \label{fig_Const_PDF_19_1pt0}
\end{subfigure}\hspace{-.1em}
\begin{subfigure}{1\textwidth}
  \centering
  \includegraphics[clip,width=.85\textwidth]{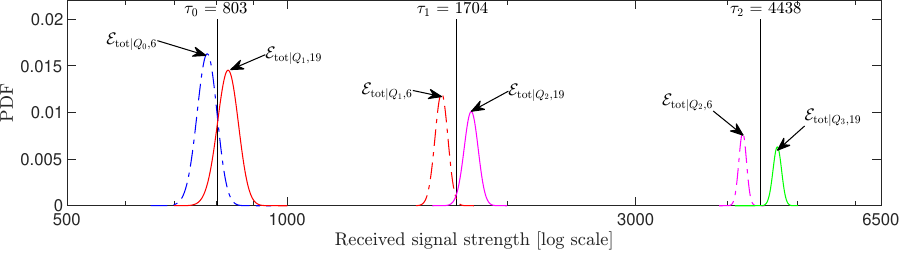}\vspace{-.1em}
  \caption{$\rho=1.12$}
  \label{fig_Const_PDF_19_1pt1}
\end{subfigure}
\caption{PDFs of the limits of received signal strength highlighting the effects of $\rho$ on the limits spacing in CSK-CT for $M=4$, $y_\text{min}=\SI{6}{\micro\metre}$, $y_\text{max}=\SI{19}{\micro\metre}$, $D=$ \SI{79.4}{\micro\metre^{2}\per\second}, $K=$ \SI{14}{}, $r=$ \SI{5}{\micro\metre}, $\Delta t=$ \SI{0.32}{\second}, and $t_\text{sym}=$ \SI{21.12}{\second}.}
\label{fig_Const_PDFvsRho}
\end{figure*}

Fig. ~\ref{fig_Const_PDFvsRho} depicts the effects of $\rho$ on the limits spacing with $M=4$. The dashed lines represent the PDFs of $\mathcal{E}_{\text{tot}|Q_\mathit{j},y_\text{min}}$ and the solid lines represent the PDFs of $\mathcal{E}_{\text{tot}|Q_{\mathit{j}+1},y_\text{max}}$. When $\rho=1$, we can observe the relation $\tau_\mathit{j} = \mathcal{E}_{\text{tot}|Q_{\mathit{j}+1},y_\text{max}}=\mathcal{E}_{\text{tot}|Q_\mathit{j},y_\text{min}} $, indicated by the overlapping between the thresholds and the limits of received signal strength. Conversely, for $\rho=1.12$, the relation $\mathcal{E}_{\text{tot}|Q_\mathit{j},y_\text{min}} <\tau_\mathit{j}<\mathcal{E}_{\text{tot}|Q_{\mathit{j}+1},y_\text{max}}$ can be observed, as indicated by the non-overlapping between the thresholds and the limits of received signal strength. This demonstrates that an increase in $\rho$ effectively widens the limits spacing. Additionally, Fig. \ref{fig_Const_PDF_19_1pt1} shows that the limits spacing widens with $\tau_j$, which implies that the errors will be the highest around $\tau_0$ and lowest around $\tau_{M-1}$.

\begin{remark}\label{remark2}
The limits spacing widens with increasing $\rho$ and $\Gamma$. Moreover, it is narrower for lower threshold indices and wider for higher indices.  $\qed$
\end{remark}

\begin{figure}[!t] 
        \centering        \includegraphics[width=.9\columnwidth,keepaspectratio]{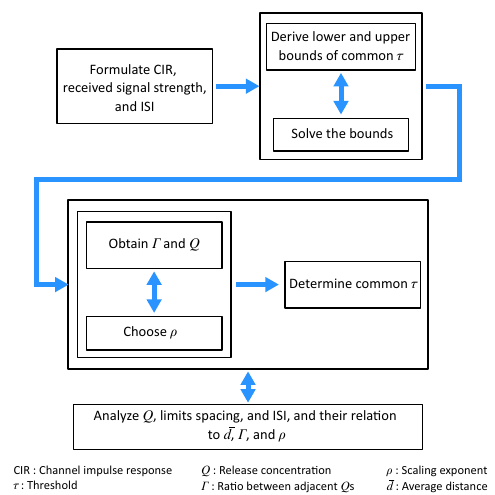}
        \caption {\label{fig_overview} Overview of the proposed CSK-CT. }
    \end{figure}

Fig. \ref{fig_overview} shows an overview of the proposed CSK-CT system.

\section{Performance Analysis}\label{sec_PerfAnaly}
\subsection{Time complexity analysis} \label{sec_timeComplex}
The operational complexity CSK imposes on a receiver is assessed herein by quantifying the rate at which the number of threshold computations grows with $M$ and $K$. To this end, the individual mathematical operations in computing a threshold are holistically perceived as a single elementary operation. Hence, the operational complexity is described by the number of thresholds computed by the receiver.

For determining the asymptotic limits of operational complexity, the time complexity metric is widely used. This standard metric estimates an algorithm's run-time by counting the number of elementary operations, typically expressed using the $\mathcal{O}$ notation \cite{Michael1996IntroToComput, Thomas2009IntoToAlgo}. The time complexity metric applies herein because computing a threshold is an elementary operation.  

\begin{figure}
\centering
  \includegraphics[width=.85\columnwidth]{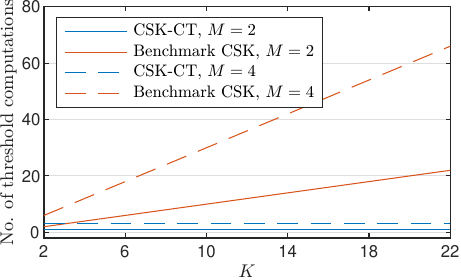}
\caption{Time complexity in terms of the number of threshold computations versus $K$. }
\label{fig_time_Complexity}
\end{figure}

From \eqref{eq_ThresholdCSK-CT}, for the proposed CSK-CT method, the number of thresholds the receiver must compute is $(M-1)$, requiring two CIRs for $y_\text{min}$ and $y_\text{max}$. This results in a time complexity of $\mathcal{O}\big(n\big) \approx \mathcal{O}(M-1)$. In contrast, for the benchmark CSK, the receiver must compute $K\cdot(M-1)$ thresholds, requiring $K$ CIRs for $y_1,y_2, \ldots, y_K$. This results in a time complexity of $\mathcal{O}\big(n^2\big) \approx \mathcal{O}(K\cdot(M-1))$. 

Fig. \ref{fig_time_Complexity} depicts the number of threshold computations for varying $K$. As expected, CSK-CT maintains a consistently low number of threshold computations, independent of $K$, resulting in $1$ and $3$ computations for BCSK and 4-CSK, respectively. In contrast, the threshold computations for the benchmark CSK increase with $K$.

\subsection{Error Probability Analysis} \label{sec_errorProb}
Considering that the symbol $S_\mathit{\!j}$ is transmitted, the receiver decodes the transmitted symbol using the following detection rule:
        \begin{equation} \label{eq_detectionrule}
                 \hat{S_\mathit{\!j}}= 
            \begin{cases}
                S_{\!0},& \text{if } {\mathcal{E}}_{\text{tot}}< \tau_0,\ \ j=0,\\
                S_\mathit{\!j},& \text{if }\tau_{\mathit{\!j}-1} \leq {\mathcal{E}}_{\text{tot}}< \tau_\mathit{j},\ \ j=1,2,\ldots,M-2,\\
                S_{\mathit{\!M}-1},&  \text{if } {\mathcal{E}}_{\text{tot}}\geq \tau_{M-2},\ \   j=M-1.
            \end{cases}
        \end{equation}
The probability of error when $S_\mathit{\!j}$ is transmitted by transmitter $k$ with distance $y_k$ is given by
    \begin{align} \label{eq_error_sj}
    P_\mathit{\!e,k}(S_\mathit{\!j}) =
    \begin{cases}
            \Pr\left({\mathcal{E}}_{\text{tot}}\geq \tau_0\right)=0.5\,\text{erfc}\left(\frac{\tau_0-\mathcal{E}_{\text{tot}|Q_0,y_k}}{\sqrt{2}\,\sigma_{\text{tot}|Q_0,y_k}}\right),\ j=0,\\
            \Pr\left({\mathcal{E}}_{\text{tot}} < \tau_{\mathit{\!j}-1} \right) + \Pr\left( {\mathcal{E}}_{\text{tot}} \ge \tau_\mathit{j}\right) \\ \ \ \ =  0.5\,\text{erfc}\left(\frac{\mathcal{E}_{\text{tot}|Q_\mathit{j},y_k}-\tau_{\mathit{\!j}-1}}{\sqrt{2}\,\sigma_{\text{tot}|Q_\mathit{j},y_k}}\right) + 0.5\,\text{erfc}\left(\frac{\tau_\mathit{j}-\mathcal{E}_{\text{tot}|Q_\mathit{j},y_k}}{\sqrt{2}\,\sigma_{\text{tot}|Q_\mathit{j},y_k}}\right),\\ \qquad \qquad j=1,2,\ldots,M-2,\\
            \Pr\left({\mathcal{E}}_{\text{tot}}<\tau_{M-2}\right)=  0.5\,\text{erfc}\left(\frac{\mathcal{E}_{\text{tot}|Q_{M-1},y_k}-\tau_{M-2}}{\sqrt{2}\,\sigma_{\text{tot}|Q_{M-1},y_k}}\right),\\ \qquad \qquad j=M-1.
        \end{cases}
    \end{align}    
From \eqref{eq_error_sj}, the probability of error for transmitter $k$, assuming each symbol is transmitted with equal probability $1/M$, is calculated as
    \begin{align} \label{eq_tot_error_prob}
        P_\mathit{\!e,k} = \frac{1}{M}\,\sum_{j=0}^{M-1}P_\mathit{\!e,k}(S_\mathit{\!j}) .
    \end{align}
Then, the total error probability is given by
    \begin{align} \label{eq_avg_tot_error_prob}
        P_\mathit{\!e}  = \frac{1}{K} \sum_{k=1}^K  P_\mathit{\!e,k}.
    \end{align}

\section{Numerical Analyses and Results}\label{sec_Results}
The transmitters are located at distances within the range of $y_\text{min}= \SI{6}{\micro\meter}$ and $y_\text{max} \in \{17, 18, 19, 20, 21\}$\,\SI{}{\micro\meter}. Without loss of generality, adjacent distances $y_k$ and $y_{k+1}$ are spaced \SI{1}{\micro\meter} apart. For example, if $y_\text{max}=\SI{17}{\micro\meter}$, then the distances are $\{y_1= \SI{6}{\micro\meter},\,y_2= \SI{7}{\micro\meter},\ldots,\, y_K= \SI{17}{\micro\meter}\}$. Accordingly, the total number of transmitters is $K = y_\text{max}-y_\text{min}+1$. The average distance is defined as $\bar{d}=(y_\text{min}+y_\text{max})/2\, \SI{}{\micro\meter}$, resulting in $\bar{d}=\{11.5, 12, 12.5, 13, 13.5\}\, \SI{}{\micro\meter}$. Table \ref{table_Sim_Param} lists the values of the remaining simulation parameters unless specified otherwise.

\begin{table}[!] 
\centering
 \caption{Simulation parameters}
\begin{tabular}{ll}
\toprule
Parameter                      & Values                   \\ \midrule
Diffusion coefficient, $D$                            & \SI{79.4}{\micro\metre^{2}\per\second} \cite{yilmaz2014Three}        \\
Radius of receiver, $r$                 & \SI{5}{\micro\metre}  \cite{yilmaz2014Three}      \\
Sampling period duration, $\Delta t$ & \SI{0.32}{\second} \cite{Meng2014OnReceiverDesign}\\
Symbol period duration, $t_\text{sym}$ & \SIrange{1.28}{32}{\second}\\ 
Min. distance between transmitters and receiver, $y_\text{min}$                          & \SI{6}{\micro\metre} \\
Max. distance between transmitters and receiver, $y_\text{max}$                          & \SIrange{17}{21}{\micro\metre} \\
Avg. distance between transmitters and receiver, $\bar{d}$                          & \SIrange{11.5}{13.5}{\micro\metre} \\
No. of transmitters, $K$                            & \SIrange{12}{16}{}             \\
Scaling exponent, $\rho$ & \SIrange{1}{4}{}\\
$Q_0$ & $1000$ molecules\\
\bottomrule                 
\end{tabular}
\label{table_Sim_Param}
\end{table}

\begin{table*}
 \centering
 \caption{Values of parameters of CSK-CT and benchmark CSK. $t_\text{sym}$ is set to $\SI{21.12}{\second}$.}
\begin{tabular}{l|lcc|ccc|ccccccc}
\toprule
\multirow{2}{*}{Scheme}                                    & \multirow{2}{*}{$\bar{d}$}                    & \multicolumn{1}{c}{\multirow{2}{*}{$\Gamma$}} & \multirow{2}{*}{$\rho$}                  & \multicolumn{3}{c|}{BCSK ($M=2$)}                                                                            & \multicolumn{7}{c}{4-CSK ($M=4$)}                                                                                                                                                                                                                               \\ 
                                                     &                                            & \multicolumn{1}{c}{}                           &                                            & \multicolumn{1}{c}{$Q_0$} & \multicolumn{1}{c}{$Q_1$} & $\tau_0$                                   & \multicolumn{1}{c}{$Q_0$} & \multicolumn{1}{c}{$Q_1$} & \multicolumn{1}{c}{$Q_2$} & \multicolumn{1}{c}{$Q_3$} & \multicolumn{1}{c}{$\tau_0$}              & \multicolumn{1}{c}{$\tau_1$}              & $\tau_2$                                   \\ \midrule
\multirow{10}{*}{CSK-CT}                           & 11.5                                      & 2.386                             &    \multirow{5}{*}{1}                                 & \multicolumn{1}{c}{1000}  & \multicolumn{1}{c}{2386}  & 478                                        & \multicolumn{1}{c}{1000}  & \multicolumn{1}{c}{2386}  & \multicolumn{1}{c}{5693}  & \multicolumn{1}{c}{13581} & \multicolumn{1}{c}{606}                   & \multicolumn{1}{c}{1184}                  & 2563                                       \\ 
                                                     & 12                                           & 2.538                                              &                                       & \multicolumn{1}{c}{1000}  & \multicolumn{1}{c}{2538}  & 483                                        & \multicolumn{1}{c}{1000}  & \multicolumn{1}{c}{2538}  & \multicolumn{1}{c}{6441}  & \multicolumn{1}{c}{16344} & \multicolumn{1}{c}{642}                   & \multicolumn{1}{c}{1283}                  & 2911                                       \\ 
                                                     & 12.5                                           & 2.695                                            &                                      & \multicolumn{1}{c}{1000}  & \multicolumn{1}{c}{2695}  & 487                                        & \multicolumn{1}{c}{1000}  & \multicolumn{1}{c}{2695}  & \multicolumn{1}{c}{7262}  & \multicolumn{1}{c}{19568} & \multicolumn{1}{c}{684}                   & \multicolumn{1}{c}{1391}                  & 3295                                       \\ 
                                                     & 13                                           & 2.857                                              &                                       & \multicolumn{1}{c}{1000}  & \multicolumn{1}{c}{2858}  & 492                                        & \multicolumn{1}{c}{1000}  & \multicolumn{1}{c}{2858}  & \multicolumn{1}{c}{8163}  & \multicolumn{1}{c}{23323} & \multicolumn{1}{c}{733}                   & \multicolumn{1}{c}{1507}                  & 3720                                       \\ 
                                                     & 13.5                                           & 3.025                                            &                                       & \multicolumn{1}{c}{1000}  & \multicolumn{1}{c}{3025}  & 496                                        & \multicolumn{1}{c}{1000}  & \multicolumn{1}{c}{3025}  & \multicolumn{1}{c}{9151}  & \multicolumn{1}{c}{27681} & \multicolumn{1}{c}{789}                   & \multicolumn{1}{c}{1633}                  & 4188                                       \\ \cmidrule{2-14} 
                                                     & 11.5                                       & 2.386                       & \multirow{5}{*}{1.24}                                      & \multicolumn{1}{c}{1000}  & \multicolumn{1}{c}{2940}  & 534                                        & \multicolumn{1}{c}{1000}  & \multicolumn{1}{c}{2940}  & \multicolumn{1}{c}{8641}  & \multicolumn{1}{c}{25400} & \multicolumn{1}{c}{780}                   & \multicolumn{1}{c}{1678}                  & 4318                                       \\ 
                                                     & 12                                              & 2.538                                        &                                  & \multicolumn{1}{c}{1000}  & \multicolumn{1}{c}{3174}  & 544                                        & \multicolumn{1}{c}{1000}  & \multicolumn{1}{c}{3174}  & \multicolumn{1}{c}{10070} & \multicolumn{1}{c}{31955} & \multicolumn{1}{c}{862}                   & \multicolumn{1}{c}{1876}                  & 5093                                       \\ 
                                                     & 12.5                                           & 2.695                                          &                & \multicolumn{1}{c}{1000}  & \multicolumn{1}{c}{3419}  & 554                                        & \multicolumn{1}{c}{1000}  & \multicolumn{1}{c}{3419}  & \multicolumn{1}{c}{11687} & \multicolumn{1}{c}{39951} & \multicolumn{1}{c}{961}                   & \multicolumn{1}{c}{2098}                  & 5982                                       \\ 
                                                     & 13                                            & 2.857                                              &                                       & \multicolumn{1}{c}{1000}  & \multicolumn{1}{c}{3676}  & 564                                        & \multicolumn{1}{c}{1000}  & \multicolumn{1}{c}{3676}  & \multicolumn{1}{c}{13511} & \multicolumn{1}{c}{49663} & \multicolumn{1}{c}{1081}                  & \multicolumn{1}{c}{2348}                  & 7001                                       \\ 
                                                     & 13.5                                           & 3.025                                            &                                       & \multicolumn{1}{c}{1000}  & \multicolumn{1}{c}{3946}  & 574                                        & \multicolumn{1}{c}{1000}  & \multicolumn{1}{c}{3946}  & \multicolumn{1}{c}{15567} & \multicolumn{1}{c}{61419} & \multicolumn{1}{c}{1224}                  & \multicolumn{1}{c}{2629}                  & 8165                                       \\ \midrule
Benchmark CSK & n.a. & n.a. & n.a. & 1000  & 1500  & n.a. & 1000  & 1500  & 2000  & 3000  & n.a. & n.a. & n.a.                  \\ \bottomrule
\end{tabular}
\label{table_Gamma_Q_tau}
\end{table*}
\begin{figure*}
\centering
\begin{subfigure}{1\textwidth}
  \centering
  \includegraphics[width=.85\textwidth]{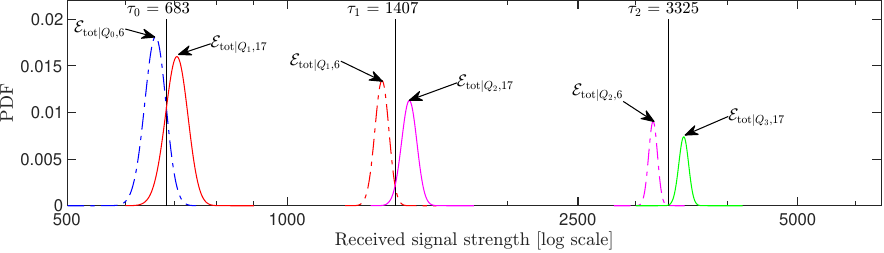}\vspace{-.1em}
  \caption{$\bar{d}=\SI{11.5}{\micro\metre}$}  \vspace{1em} \label{fig_Const_PDF_17_1pt1}
\end{subfigure}
\begin{subfigure}{1\textwidth}
  \centering
  \includegraphics[width=.85\textwidth]{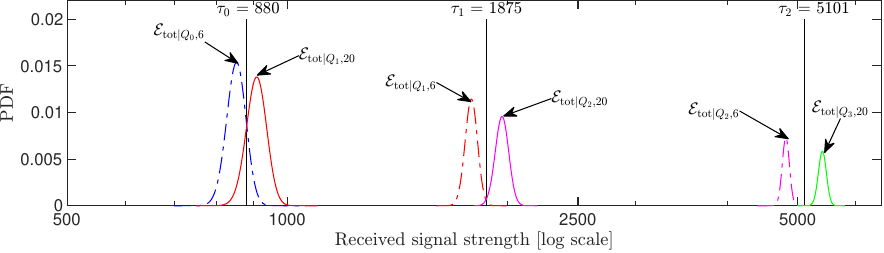}\vspace{-.1em}
  \caption{$\bar{d}=\SI{13}{\micro\metre}$}    \label{fig_Const_PDF_20_1pt1}
\end{subfigure}
\caption{PDFs of the limits of received signal strength highlighting the effects of $\bar{d}$ on the limits spacing in CSK-CT for $M=4$, $\rho=1.12$, and $t_\text{sym}=$ \SI{21.12}{\second}. }
\label{fig_Const_PDFvsDistance}
\end{figure*}

In Table \ref{table_Gamma_Q_tau}, we present the $Q$ values for both CSK-CT and benchmark CSK and the $\Gamma$ and $\tau$ values of CSK-CT corresponding to varying $\bar{d}$ and $\rho$. For CSK-CT, the values of $\Gamma$, $Q$, and $\tau$ are obtained using \eqref{eq_Gamma_Final}, \eqref{eq_DetQInequal}, and \eqref{eq_ThresholdCSK-CT}, respectively. As $\bar{d}$ increases, $\Gamma$ increases, which causes $Q$ and $\tau$ to increase. An increase in $\rho$ also causes $Q$ and $\tau$ to increase because it has an effect similar to increasing $\Gamma$. While the increase in $Q$ is attributed to the relation to $\Gamma$ and $\rho$, the increase in $\tau$ is because the received signal strength increases with $Q$. $\tau_0$ is larger in 4-CSK than in BCSK, despite identical $Q_0$ and $Q_1$ values. This is due to the higher average release concentration in 4-CSK, primarily influenced by significantly higher $Q_2$ and $Q_3$ values.

Table \ref{table_Gamma_Q_tau} also shows that CSK-CT generally exhibits higher $Q$ values compared to the benchmark CSK because its $Q$ values are distance-dependent, designed to ensure the thresholds accommodate the weakest received signal strength (see \eqref{eq_ini_ConditionCSK-CT}). Typical values of $Q$ range between $1,000$ and $100,000$ molecules \cite{Mahfuz2014Comprehensive, Singhal2015Performance, Jamali2018Noncoherent, Okaie2020MobileMolCom}. However, a larger $Q$ value implies a corresponding increase in the transmission energy \cite{Kuran2010Energy,Kuscu2019TxRxArchitectures}. The optimization of $Q$ values considering the energy budget of a bio-nanothing is left aside for future work. The $Q$ values of benchmark CSK are distance-independent and are comparable to those used in \cite{Singhal2015Performance}.

\begin{figure*}[t!]
\centering
\begin{subfigure}{.46\textwidth}
  \centering
        \includegraphics[width=.85\columnwidth,keepaspectratio]{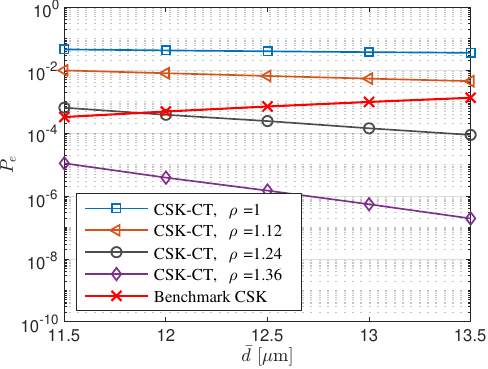}
        \caption{BCSK}
        \label{fig_error_prob_BCSK}
\end{subfigure}\hspace{1em}%
\begin{subfigure}{.46\textwidth}
  \centering
        \includegraphics[width=.85\columnwidth,keepaspectratio]{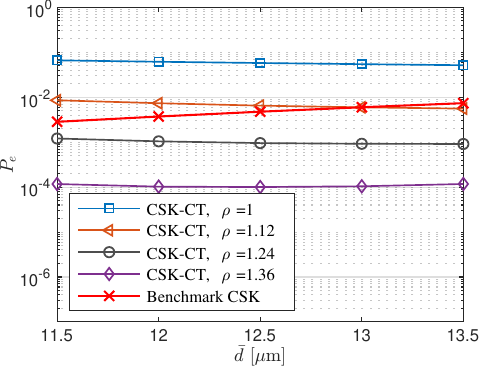}
       \caption{4-CSK}
    \label{fig_error_prob_4CSK}
\end{subfigure}\hspace{1em}%
    \caption{$P_\mathit{\!e}$ versus $\bar{d}$ and varying $\rho$ for $t_\text{sym}=\SI{21.12}{\second}$. }
\label{fig_Error_probability}
\end{figure*} 

\begin{figure*}
\centering
\begin{subfigure}{.46\textwidth}
  \centering
        \includegraphics[width=1\columnwidth,keepaspectratio]{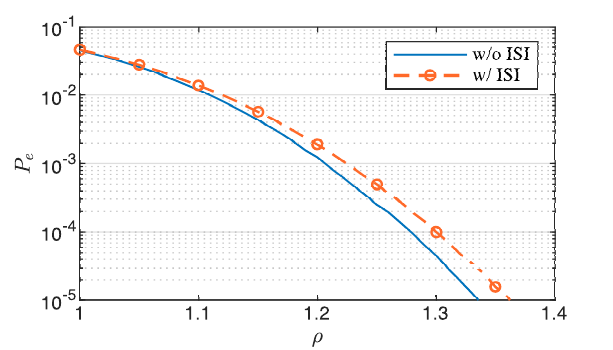}
        \caption{BCSK}
        \label{fig_error_VS_rho_BCSK}
\end{subfigure}\hspace{1em}%
\begin{subfigure}{.46\textwidth}
  \centering
        \includegraphics[width=.9\columnwidth,keepaspectratio]{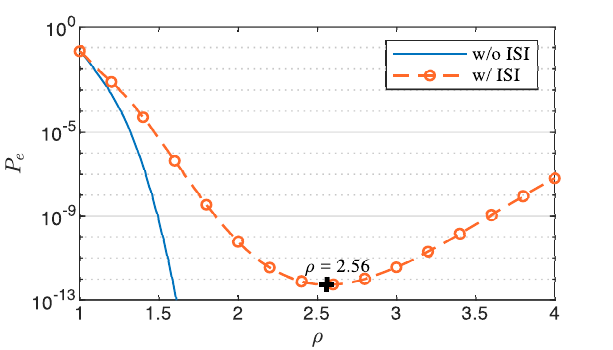}
       \caption{4-CSK}
    \label{fig_error_VS_rho_4CSK}
\end{subfigure}\hspace{1em}%
    \caption{$P_\mathit{\!e}$ versus $\rho$ and varying $\rho$ for $t_\text{sym}=\SI{21.12}{\second}$ and $\bar{d}=\SI{11.5}{\micro\meter}$. The $x$- and $y$-axis for BCSK is cropped for better visualization, and the black cross in 4-CSK indicates the optimal value of $\rho$.}
\label{fig_error_VS_rho_COMB}
\end{figure*}

\begin{figure}
\centering
        \includegraphics[width=.9\columnwidth,keepaspectratio]{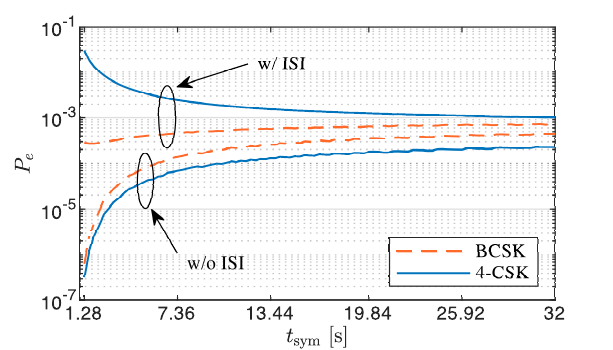}
        \caption{$P_\mathit{\!e}$ versus $t_\text{sym}$ for $\rho=1.24$ and $\bar{d}=\SI{11.5}{\micro\meter}$.}
        \label{fig_Error_Vs_tsym}
\end{figure}

Fig. \ref{fig_Const_PDFvsDistance} shows the impact of $\bar{d}$ on the limits spacing. The limits spacing widens with $\bar{d}$ because $Q$ increases with $\bar{d}$. As $Q$ increases, the separation between the limits of received signal strength increases proportionally, resulting in wider limits spacing. An increase in $Q$ is not only beneficial (wider limit spacing) to $P_\mathit{\!e}$, but it can also be detrimental (increased ISI). Widening of limits spacing is referred to as the \textit{positive effect} and increases in ISI as the \textit{negative effect} on $P_\mathit{\!e}$.

Fig. \ref{fig_Error_probability} presents the behavior of the $P_\mathit{\!e}$ for varying $\bar{d}$. Generally, $P_\mathit{\!e}$ improves as $\bar{d}$ increases due to the dominance of the positive effect over the negative effect. For $\rho$ = 1.24 and 1.36 cases of 4-CSK, $P_\mathit{\!e}$ degrades with $\bar{d}$ as the negative effect dominates the positive effect. The findings indicate that CSK-CT can be strategically designed to mitigate ISI. For the benchmark CSK, the performance degrades because the release concentration is not determined considering $\bar{d}$. CSK-CT outperforms the benchmark CSK for larger $\rho$. For example, with $\rho=1.36$, CSK-CT achieves $P_\mathit{\!e}$ up to $10^{-7}$ and $10^{-4}$ for BCSK and 4-CSK, respectively, which is significantly lower than the benchmark CSK's performance of $10^{-4}$ and $10^{-3}$ for the same $M$ values. Further performance enhancements are attainable by increasing $\rho$ (see Fig. \ref{fig_error_VS_rho_COMB}). This demonstrates CSK-CT's simplicity in achieving a desired $P_\mathit{\!e}$ by selecting an appropriate $\rho$ value, compared to the potentially exhaustive optimization required in benchmark CSK.

Fig. \ref{fig_error_VS_rho_COMB} shows the behavior of $P_\mathit{\!e}$ for varying $\rho$. In the absence of ISI, $P_\mathit{\!e}$ improves with $\rho$ due to wider limits spacing, as indicated by Remark \ref{remark2}. However, in the presence of ISI, $P_\mathit{\!e}$ of BCSK improves with $\rho$ as the positive effect dominates the negative effect. On the other hand,  $P_\mathit{\!e}$ of 4-CSK improves until the optimal value ($\rho=2.56$, in this case), beyond which it begins to worsen. This indicates that the positive effect remains dominant over the negative effect until the optimal value, after which the negative effect dominates the positive effect.

Fig. \ref{fig_Error_Vs_tsym} shows the behavior of $P_\mathit{\!e}$ for varying $t_\text{sym}$. In the absence of ISI, $P_\mathit{\!e}$ increases with $t_\text{sym}$ because $\Gamma$ reduces with $t_\text{sym}$, as stated in Remark \ref{remark1}. A lower $\Gamma$ results in closer $Q$ values (see Table \ref{table_Gamma_Q_tau}), leading to narrow limits spacing. As $t_\text{sym}$ further increases, the rise in $P_\mathit{\!e}$ becomes marginal because the received signal strength converges to a steady state \cite{Wang2019ANovel}, resulting in incremental narrowing of the limits spacing. Interestingly, 4-CSK performs better than BCSK because its $P_\mathit{\!e}$ depends not only on the high errors around $\tau_0$ but also on the low errors around $\tau_1$ and $\tau_2$, where the limits spacings are significantly wider (see Remark \ref{remark2}). In contrast, BCSK's $P_\mathit{\!e}$ depends solely on high errors around $\tau_0$. 

In the presence of ISI, 4-CSK's $P_\mathit{\!e}$ is higher than that of BCSK, as expected. 4-CSK exhibits poor performance for smaller $t_\text{sym}$ values because there is no assurance that the received signal strength is much larger than the ISI at shorter symbol periods. As $t_\text{sym}$ and the received signal strength increase, the $P_\mathit{\!e}$ of 4-CSK gradually improves and stabilizes, reflecting the converging property of the received signal strength. Conversely, BCSK exhibits a slight increase in $P_\mathit{\!e}$ with increasing $t_\text{sym}$ due to the effects of the narrowing limits spacing.

\section{Conclusion and Discussions} \label{sec_conclusions}
This study proposed CSK-CT as a solution to the high complexity issue prevalent in existing CSK methods. By leveraging the fact that the received signal strength is distance-dependent, CSK-CT uses common thresholds and achieves a time complexity of $\mathcal{O}\big(n\big)$, offering a significant improvement over the $\mathcal{O}\big(n^2\big)$ time complexity of the benchmark CSK. Additionally, CSK-CT attains lower error probabilities than the benchmark CSK when the scaling exponent value is appropriately selected. Key findings include identifying an optimum scaling exponent value for 4-CSK and increasing error probabilities in BCSK for longer symbol period durations. CSK-CT requires two CIRs and has ISI combating ability, an advantage over the benchmark CSK.

Owing to its lower time complexity and error probabilities, CSK-CT is particularly advantageous for nano-sensor networks in MC-based applications focused on data-gathering tasks such as disease bio-marker detection in the bloodstream and toxic chemical detection in industrial settings. These applications often involve deploying multiple nano-sensors that sense, collect, and report data to a nano-sink, which typically operates under computational and energy constraints.

CSK-CT requires larger release concentration values, implicating increased transmission energy costs besides ISI. Studies have used release concentrations similar to those used in CSK-CT \cite{Mahfuz2014Comprehensive, Jamali2018Noncoherent, Okaie2020MobileMolCom}, but balancing the trade-off between error probability and transmission energy cost is critical. Future research directions include optimizing release concentrations and enhancing CSK-CT's energy efficiency and robustness against ISI for demanding nano-sensor networks with higher reliability requirements. Investigating CSK-CT under time-varying channel conditions, such as mobile nano-sensors and flow-assisted diffusion, is also established as a future research direction.

\bibliographystyle{IEEEtran}
\bibliography{Refs_CSK-CT} 

\begin{thebibliography}{10}
\providecommand{\url}[1]{#1}
\csname url@samestyle\endcsname
\providecommand{\newblock}{\relax}
\providecommand{\bibinfo}[2]{#2}
\providecommand{\BIBentrySTDinterwordspacing}{\spaceskip=0pt\relax}
\providecommand{\BIBentryALTinterwordstretchfactor}{4}
\providecommand{\BIBentryALTinterwordspacing}{\spaceskip=\fontdimen2\font plus
\BIBentryALTinterwordstretchfactor\fontdimen3\font minus
  \fontdimen4\font\relax}
\providecommand{\BIBforeignlanguage}[2]{{%
\expandafter\ifx\csname l@#1\endcsname\relax
\typeout{** WARNING: IEEEtran.bst: No hyphenation pattern has been}%
\typeout{** loaded for the language `#1'. Using the pattern for}%
\typeout{** the default language instead.}%
\else
\language=\csname l@#1\endcsname
\fi
#2}}
\providecommand{\BIBdecl}{\relax}
\BIBdecl

\bibitem{Akyildiz2015IoBnT}
I.~F. Akyildiz, M.~Pierobon, S.~Balasubramaniam, and Y.~Koucheryavy, ``{The
  Internet of Bio-Nano things},'' \emph{IEEE Communications Magazine}, vol.~53,
  no.~3, pp. 32--40, 2015.

\bibitem{Li2022IoNTInterface}
Y.~Li, L.~Lin, W.~Guo, D.~Zhang, and K.~Yang, ``Error performance and mutual
  information for iont interface system,'' \emph{IEEE Internet of Things
  Journal}, vol.~9, no.~12, pp. 9831--9842, 2022.

\bibitem{akyildiz2008nanonetworks}
I.~F. Akyildiz, F.~Brunetti, and C.~Bl\'{a}zquez, ``{Nanonetworks: A New
  Communication Paradigm},'' \emph{Elsevier Computer Networks}, vol.~52,
  no.~12, pp. 2260--2279, Aug. 2008.

\bibitem{Felicetti2016Applications}
\BIBentryALTinterwordspacing
L.~Felicetti, M.~Femminella, G.~Reali, and P.~Liò, ``{Applications of
  Molecular Communications to Medicine: A Survey},'' \emph{Nano Communication
  Networks}, vol.~7, pp. 27--45, 2016. [Online]. Available:
  \url{https://www.sciencedirect.com/science/article/pii/S1878778915000411}
\BIBentrySTDinterwordspacing

\bibitem{Akkas2020Healthcare}
\BIBentryALTinterwordspacing
M.~Akkaş, R.~Sokullu, and H.~{Ertürk Çetin}, ``{Healthcare and Patient
  Monitoring using IoT},'' \emph{Elsevier Internet of Things}, vol.~11, p.
  100173, 2020. [Online]. Available:
  \url{https://www.sciencedirect.com/science/article/pii/S2542660520300147}
\BIBentrySTDinterwordspacing

\bibitem{Shitiri2021Timing}
E.~Shitiri and H.-S. Cho, ``{Timing Alignment in Molecular-Communication-Based
  Nanonetworks},'' \emph{IEEE Communications Magazine}, vol.~59, no.~5, pp.
  54--60, 2021.

\bibitem{Al-Zubi2022Intrabody}
M.~M. Al-Zubi, A.~S. Mohan, P.~Plapper, and S.~H. Ling, ``{Intrabody Molecular
  Communication via Blood-Tissue Barrier for Internet of Bio-Nano Things},''
  \emph{IEEE Internet of Things Journal}, vol.~9, no.~21, pp. 21\,802--21\,810,
  2022.

\bibitem{nakano2005molecular}
Y.~Moritani, S.~Hiyama, T.~Suda, R.~Egashira, A.~Enomoto, M.~Moore, and
  T.~Nakano, ``{Molecular Communications between Nanomachines},'' in \emph{24th
  IEEE Conference on Computer Communications (IEEE INFOCOM 2005)}, March 2005.

\bibitem{Chude2017SurveyMCTDDD}
U.~A.~K. {Chude-Okonkwo}, R.~{Malekian}, B.~T. {Maharaj}, and A.~V.
  {Vasilakos}, ``{Molecular Communication and Nanonetwork for Targeted Drug
  Delivery: A Survey},'' \emph{IEEE Communications Surveys and Tutorials},
  vol.~19, no.~4, pp. 3046--3096, 2017.

\bibitem{Guo2021}
W.~Guo, M.~Abbaszadeh, L.~Lin, J.~Charmet, P.~Thomas, Z.~Wei, B.~Li, and
  C.~Zhao, ``{Molecular Physical Layer for 6G in Wave-Denied Environments},''
  \emph{IEEE Communications Magazine}, vol.~59, no.~5, pp. 33--39, 2021.

\bibitem{Veletic2022Omplants}
M.~Veletić, E.~H. Apu, M.~Simić, J.~Bergsland, I.~Balasingham, C.~H. Contag,
  and N.~Ashammakhi, ``Implants with sensing capabilities,'' \emph{Chemical
  Reviews}, vol. 122, no.~21, pp. 16\,329--16\,363, 2022, pMID: 35981266.

\bibitem{Reza2019EarlyCancer}
R.~Mosayebi, A.~Ahmadzadeh, W.~Wicke, V.~Jamali, R.~Schober, and
  M.~Nasiri-Kenari, ``Early cancer detection in blood vessels using mobile
  nanosensors,'' \emph{IEEE Transactions on NanoBioscience}, vol.~18, no.~2,
  pp. 103--116, 2019.

\bibitem{Akyl2020Panacea}
I.~F. Akyildiz, M.~Ghovanloo, U.~Guler, T.~Ozkaya-Ahmadov, A.~F. Sarioglu, and
  B.~D. Unluturk, ``Panacea: An internet of bio-nanothings application for
  early detection and mitigation of infectious diseases,'' \emph{IEEE Access},
  vol.~8, pp. 140\,512--140\,523, 2020.

\bibitem{Mahfuz2010characterization}
\BIBentryALTinterwordspacing
M.~U. Mahfuz, D.~Makrakis, and H.~T. Mouftah, ``{On The Characterization of
  Binary Concentration-Encoded Molecular Communication in Nanonetworks},''
  \emph{Elesevier Nano Communication Networks}, vol.~1, no.~4, pp. 289--300,
  2010. [Online]. Available:
  \url{https://www.sciencedirect.com/science/article/pii/S1878778911000020}
\BIBentrySTDinterwordspacing

\bibitem{Mahfuz2014Comprehensive}
------, ``{A Comprehensive Study of Sampling-Based Optimum Signal Detection in
  Concentration-Encoded Molecular Communication},'' \emph{IEEE Transactions on
  NanoBioscience}, vol.~13, no.~3, pp. 208--222, 2014.

\bibitem{Singhal2015Performance}
A.~Singhal, R.~K. Mallik, and B.~Lall, ``{Performance Analysis of Amplitude
  Modulation Schemes for Diffusion-Based Molecular Communication},'' \emph{IEEE
  Transactions on Wireless Communications}, vol.~14, no.~10, pp. 5681--5691,
  2015.

\bibitem{Jamali2018Noncoherent}
V.~Jamali, N.~Farsad, R.~Schober, and A.~Goldsmith, ``{Non-Coherent Detection
  for Diffusive Molecular Communication Systems},'' \emph{IEEE Transactions on
  Communications}, vol.~66, no.~6, pp. 2515--2531, 2018.

\bibitem{Wang2023EffConsteCSK}
C.~Wang, X.~Chen, Y.~Tang, B.~Li, Y.~Huang, D.~Tang, and M.~Wen, ``{An
  Effective Constellation Design for Concentration Shift Keying in Molecular
  Communication Systems},'' \emph{IEEE Internet of Things Journal}, pp. 1--1,
  2023.

\bibitem{Kim2012Isomers}
N.-R. Kim and C.-B. Chae, ``{Novel Modulation Techniques Using Isomers as
  Messenger Molecules for Molecular Communication Via Diffusion},'' in
  \emph{2012 IEEE International Conference on Communications (ICC)}, 2012, pp.
  6146--6150.

\bibitem{Kabir2015DMoSK}
M.~H. Kabir, S.~M. Riazul~Islam, and K.~S. Kwak, ``{D-MoSK Modulation in
  Molecular Communications},'' \emph{IEEE Transactions on NanoBioscience},
  vol.~14, no.~6, pp. 680--683, 2015.

\bibitem{Wang2020PerforAnalyD-MoSK}
J.~Wang, X.~Liu, M.~Peng, and M.~Daneshmand, ``{Performance Analysis of D-MoSK
  Modulation in Mobile Diffusive-Drift Molecular Communications},'' \emph{IEEE
  Internet of Things Journal}, vol.~7, no.~11, pp. 11\,318--11\,326, 2020.

\bibitem{Tang2021MolTypePerm}
Y.~Tang, Y.~Huang, C.-B. Chae, W.~Duan, M.~Wen, and L.-L. Yang,
  ``Molecular-type permutation shift keying in molecular mimo communications
  for iobnt,'' \emph{IEEE Internet of Things Journal}, vol.~8, no.~21, pp.
  16\,023--16\,034, 2021.

\bibitem{Srinivas2012TimeMod}
K.~V. Srinivas, A.~W. Eckford, and R.~S. Adve, ``{Molecular Communication in
  Fluid Media: The Additive Inverse Gaussian Noise Channel},'' \emph{IEEE
  Transactions on Information Theory}, vol.~58, no.~7, pp. 4678--4692, 2012.

\bibitem{Aeeneh2020TimingModulation}
S.~Aeeneh, N.~Zlatanov, A.~Gohari, M.~Nasiri-Kenari, and M.~Mirmohseni,
  ``{Timing Modulation for Macro-Scale Molecular Communication},'' \emph{IEEE
  Wireless Communications Letters}, vol.~9, no.~9, pp. 1356--1360, 2020.

\bibitem{Li2020Time-BasedModulation}
Q.~Li, ``{A Novel Time-Based Modulation Scheme in Time-Asynchronous Channels
  for Molecular Communications},'' \emph{IEEE Transactions on NanoBioscience},
  vol.~19, no.~1, pp. 59--67, 2020.

\bibitem{Aghababaiyan2020DSK}
K.~Aghababaiyan, V.~Shah-Mansouri, and B.~Maham, ``{Direction Shift Keying
  Modulation for Molecular Communication},'' in \emph{IEEE International
  Conference on Communications (ICC)}, 2020, pp. 1--6.

\bibitem{Aghababaiyan2022BDSK}
K.~Aghababaiyan, H.~Kebriaei, V.~Shah-Mansouri, B.~Maham, and D.~Niyato,
  ``{Enhanced Modulation for Multiuser Molecular Communication in Internet of
  Nano Things},'' \emph{IEEE Internet of Things Journal}, vol.~9, no.~20, pp.
  19\,787--19\,802, 2022.

\bibitem{Brand2023SwitchingMolecules}
L.~Brand, M.~Scherer, S.~Lotter, T.~t. Dieck, M.~Schaufer, A.~Burkovski,
  H.~Sticht, K.~Castiglione, and R.~Schober, ``{Switchable Signaling Molecules
  for Media Modulation: Fundamentals, Applications, and Research Directions},''
  \emph{IEEE Communications Magazine}, pp. 1--7, 2023.

\bibitem{Song2019DNAlogiccircuits}
T.~Song, A.~Eshra, S.~Shah, H.~Bui, D.~Fu, M.~Yang, R.~Mokhtar, and J.~Reif,
  ``Fast and compact dna logic circuits based on single-stranded gates using
  strand-displacing polymerase,'' \emph{Nature Nanotechnology}, Nov 2019.

\bibitem{Wang2020DNAswitchingcircuits}
\BIBentryALTinterwordspacing
F.~Wang, H.~Lv, Q.~Li, J.~Li, X.~Zhang, J.~Shi, L.~Wang, and C.~Fan,
  ``Implementing digital computing with dna-based switching circuits,''
  \emph{Nature Communications}, vol.~11, no.~1, p. 121, Jan 2020. [Online].
  Available: \url{https://doi.org/10.1038/s41467-019-13980-y}
\BIBentrySTDinterwordspacing

\bibitem{Liu2022DNAComputingMC}
Q.~Liu, K.~Yang, J.~Xie, and Y.~Sun, ``Dna-based molecular computing, storage,
  and communications,'' \emph{IEEE Internet of Things Journal}, vol.~9, no.~2,
  pp. 897--915, 2022.

\bibitem{jamali2019channelTut}
V.~Jamali, A.~Ahmadzadeh, W.~Wicke, A.~Noel, and R.~Schober, ``{Channel
  Modeling for Diffusive Molecular Communication—A Tutorial Review},''
  \emph{Proceedings of the IEEE}, vol. 107, no.~7, pp. 1256--1301, 2019.

\bibitem{Kuran2021SurveyMod}
M.~{\c{S}}. Kuran, H.~B. Yilmaz, I.~Demirkol, N.~Farsad, and A.~Goldsmith, ``{A
  Survey on Modulation Techniques in Molecular Communication via Diffusion},''
  \emph{{IEEE Communications Surveys \& Tutorials}}, vol.~23, no.~1, pp. 7--28,
  2021.

\bibitem{Mahfuz2011ISI}
M.~U. Mahfuz, D.~Makrakis, and H.~T. Mouftah, ``{Characterization of
  intersymbol interference in concentration-encoded unicast molecular
  communication},'' in \emph{2011 24th Canadian Conference on Electrical and
  Computer Engineering(CCECE)}, 2011, pp. 000\,164--000\,168.

\bibitem{Lin2016ClockSync}
L.~Lin, C.~Yang, M.~Ma, S.~Ma, and H.~Yan, ``A clock synchronization method for
  molecular nanomachines in bionanosensor networks,'' \emph{IEEE Sensors
  Journal}, vol.~16, no.~19, pp. 7194--7203, 2016.

\bibitem{Shitiri2021TDMA}
E.~Shitiri and H.-S. Cho, ``{A TDMA-Based Data Gathering Protocol for Molecular
  Communication via Diffusion-Based Nano-Sensor Networks},'' \emph{IEEE Sensors
  Journal}, vol.~21, no.~17, pp. 19\,582--19\,595, 2021.

\bibitem{einstein1956BM}
A.~Einstein, \emph{{Investigations on the Theory of the Brownian
  movement}}.\hskip 1em plus 0.5em minus 0.4em\relax Dover Publications, 1956.

\bibitem{yilmaz2014Three}
H.~B. {Yilmaz}, A.~C. {Heren}, T.~{Tugcu}, and C.~{Chae}, ``{Three-Dimensional
  Channel Characteristics for Molecular Communications With an Absorbing
  Receiver},'' \emph{IEEE Communications Letters}, vol.~18, no.~6, pp.
  929--932, June 2014.

\bibitem{Kuscu2019TxRxArchitectures}
M.~Kuscu, E.~Dinc, B.~A. Bilgin, H.~Ramezani, and O.~B. Akan, ``{Transmitter
  and Receiver Architectures for Molecular Communications: A Survey on Physical
  Design With Modulation, Coding, and Detection Techniques},''
  \emph{Proceedings of the IEEE}, vol. 107, no.~7, pp. 1302--1341, 2019.

\bibitem{redner2001AGuide}
S.~Redner, \emph{{A Guide to First-Passage Processes}}.\hskip 1em plus 0.5em
  minus 0.4em\relax Cambridge University Press, 2001.

\bibitem{Kuran2010Energy}
\BIBentryALTinterwordspacing
M.~Şükrü Kuran, H.~B. Yilmaz, T.~Tugcu, and B.~Özerman, ``{Energy Model for
  Communication via Diffusion in Nanonetworks},'' \emph{Elsevier Nano
  Communication Networks}, vol.~1, no.~2, pp. 86--95, 2010. [Online].
  Available:
  \url{https://www.sciencedirect.com/science/article/pii/S1878778910000219}
\BIBentrySTDinterwordspacing

\bibitem{Hoda2013BlindSynchronization}
H.~ShahMohammadian, G.~G. Messier, and S.~Magierowski, ``{Blind Synchronization
  in Diffusion-Based Molecular Communication Channels},'' \emph{IEEE
  Communications Letters}, vol.~17, no.~11, pp. 2156--2159, 2013.

\bibitem{Wang2019ANovel}
Y.~Wang, A.~Noel, and N.~Yang, ``{A Novel $A~Priori$ Simulation Algorithm for
  Absorbing Receivers in Diffusion-Based Molecular Communication Systems},''
  \emph{IEEE Transactions on NanoBioscience}, vol.~18, no.~3, pp. 437--447,
  2019.

\bibitem{Michael1996IntroToComput}
M.~Sipser, \emph{Introduction to the Theory of Computation}, 1st~ed.\hskip 1em
  plus 0.5em minus 0.4em\relax International Thomson Publishing, 1996.

\bibitem{Thomas2009IntoToAlgo}
T.~H. Cormen, C.~E. Leiserson, R.~L. Rivest, and C.~Stein, \emph{Introduction
  to Algorithms, Third Edition}, 3rd~ed.\hskip 1em plus 0.5em minus 0.4em\relax
  The MIT Press, 2009.

\bibitem{Meng2014OnReceiverDesign}
L.-S. Meng, P.-C. Yeh, K.-C. Chen, and I.~F. Akyildiz, ``{On Receiver Design
  for Diffusion-Based Molecular Communication},'' \emph{IEEE Transactions on
  Signal Processing}, vol.~62, no.~22, pp. 6032--6044, 2014.

\bibitem{Okaie2020MobileMolCom}
Y.~Okaie and T.~Nakano, ``Mobile molecular communication through multiple
  measurements of the concentration of molecules,'' \emph{IEEE Access}, vol.~8,
  pp. 179\,606--179\,615, 2020.

\end{thebibliography}

\begin{IEEEbiography}[{\includegraphics[width=1in,height=1.25in,clip,keepaspectratio]{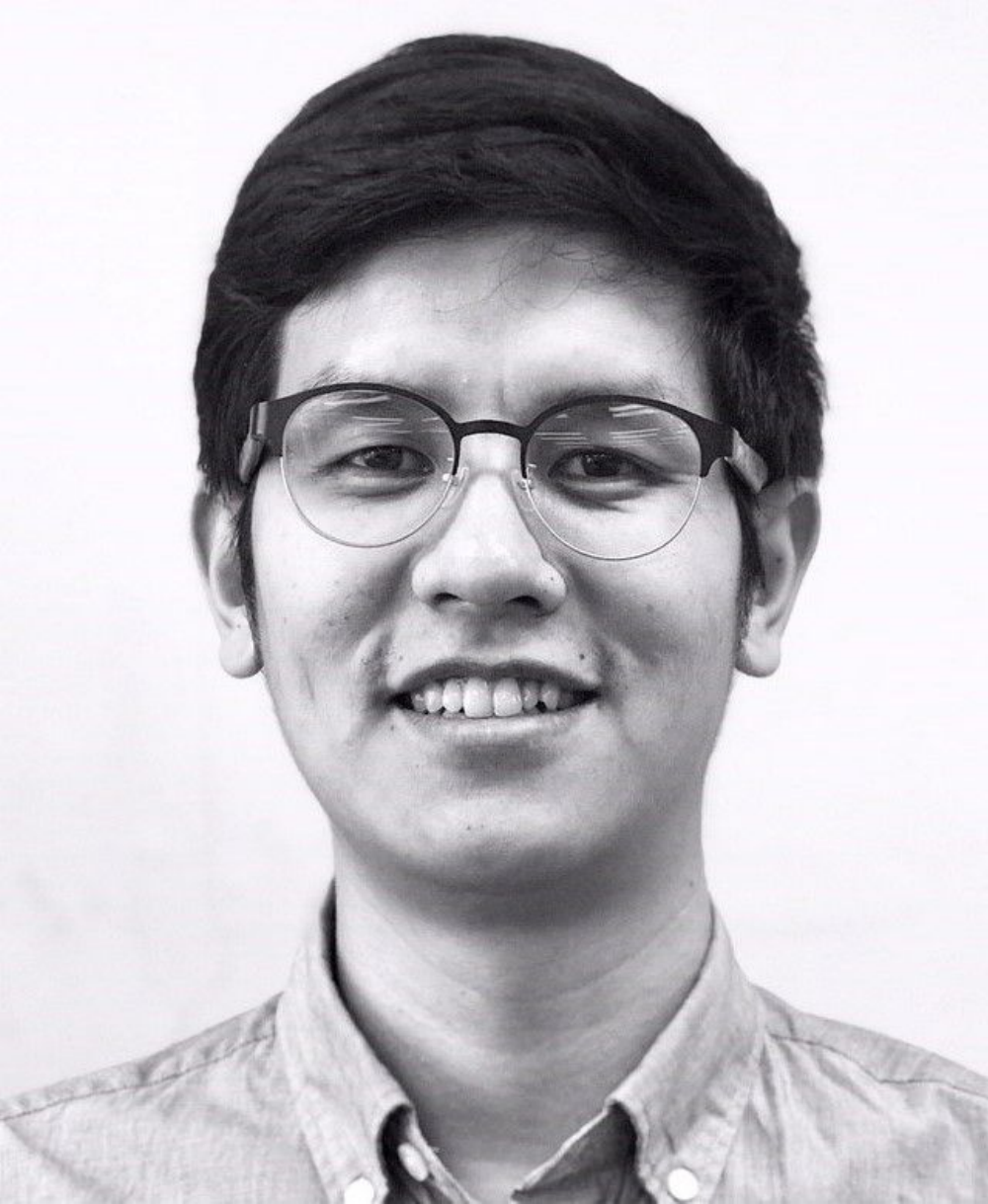}}]{Ethungshan Shitiri}
(Member, IEEE) received the B.E. degree in electronics and communication engineering (2006), the M.Tech. degree in communication systems (2013), and the Ph.D. degree in information and communication engineering (2018) from Thanthai Periyar Government Institute of Technology (India), Christ University (India), and Kyungpook National University (KNU) (South Korea), respectively. He was a Postdoctoral Research Fellow with the IDEC Center, KNU (2018-2021) and the Electronics and Electrical Engineering Advancement Institute, KNU (2021-2023). Since 2023, he has been with N3Cat at Universitat Politècnica de Catalunya, working on intra-body nanoscale communication systems. His research interests include resource allocation, handshaking protocols, synchronization, and multiple access techniques for wireless communication networks—namely cellular networks, M2M networks, underwater acoustic sensor networks, and molecular communication networks. He was a recipient of the KNU Honors Scholarship (2014), Qualcomm Innovation Award (2016), Best Thesis Award of School of Electronics (2018), and Best Paper Award, KICS (2021).
\end{IEEEbiography}

\begin{IEEEbiography}[{\includegraphics[width=1in,height=1.25in,clip,keepaspectratio]{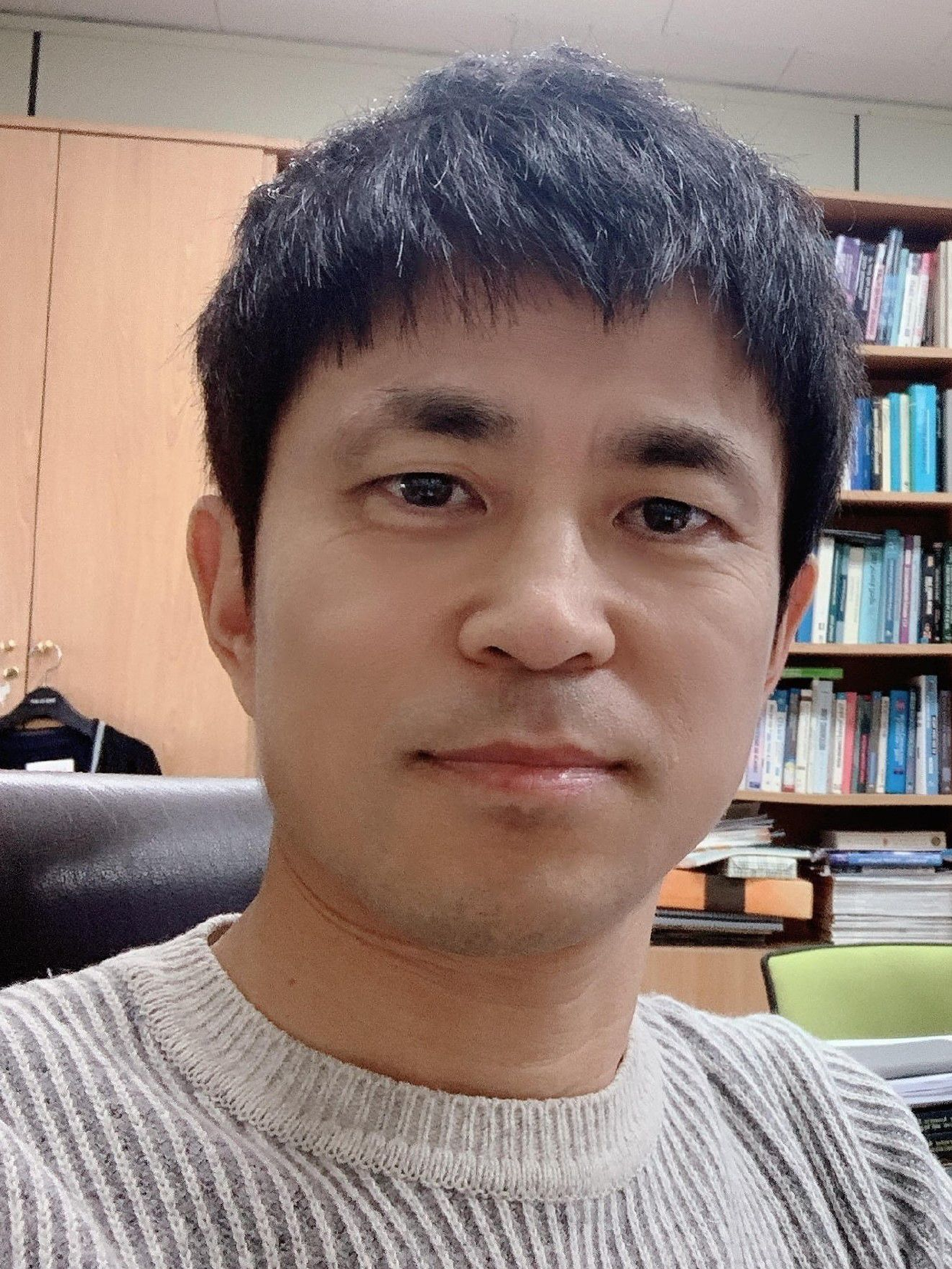}}]{Ho-Shin Cho}(Senior Member, IEEE) received the B.S., M.S., and Ph.D. degrees in electrical engineering from the Korea Advanced Institute of Science and Technology (KAIST), in 1992, 1994, and 1999, respectively. From 1999 to 2000, he was a Senior Member of the Research Staff with the Electronics and Telecommunications Research Institute (ETRI), where he was involved in developing a base station system for IMT-2000. From 2001 to 2002, he was a Faculty Member with the School of Electronics, Telecommunications, and Computer Engineering, Hankuk Aviation University. In 2003, he joined the School of Electronics Engineering, Kyungpook National University, as a Faculty Member, where he is currently a Professor. His research interests include traffic engineering, radio resource management, and medium access control protocol for wireless communication networks, cellular networks, underwater communication, and molecular communication.
Prof. Cho is a member of IEICE, IEEK, KICS, and ASK. He received the Rising Researcher Fellowship of NRF, in 1998.

\end{IEEEbiography}

\end{document}